\documentclass[aps,prl,reprint,groupedaddress,amsmath,amssymb]{revtex4-1}

\usepackage{graphicx,color}
\usepackage{dcolumn,soul}

\begin{document}


\title{A probability distribution for quantum tunneling times}


\author{Jos\'e T. Lunardi}
\email[]{jttlunardi@uepg.br}
\affiliation{Department of Mathematics \& Statistics, State University of Ponta Grossa, Avenida Carlos Cavalcanti 4748, Cep 84030-900, Ponta Grossa, PR, Brazil}

\author{Luiz A. Manzoni}
\email[]{manzoni@cord.edu}
\affiliation{Department of Physics, Concordia
College, 901 8th St. S., Moorhead, MN 56562, USA}


\date{\today}

\begin{abstract}
We propose a general expression for the probability distribution of real-valued tunneling times of a localized particle, as measured by the Salecker-Wigner-Peres quantum clock. This general expression is used to obtain the distribution of times for the scattering of a particle through a static rectangular barrier and for the tunneling decay of an initially bound state after the sudden deformation of the potential, the latter case being relevant to understand tunneling times in recent attosecond experiments involving strong field ionization.
\end{abstract}

\pacs{}

\maketitle

\section{\label{intro}Introduction}

The search for a proper definition of quantum tunneling times for massive particles, having well-behaved properties for a wide range of parameters, has remained an important and open \emph{theoretical} problem since, essentially, the inception of quantum mechanics (see, e.g., \cite{Win06, Win06a} and references therein). However, such tunneling times were beyond the experimental reach until recent advances in ultrafast physics have made possible measurements of time in the attosecond scale, opening up the experimental possibility of measuring electronic tunneling times through a classically forbidden region \cite{LWM14, TMK15, POS15, ZMD16} and reigniting the discussion of tunneling times. Still, the intrinsic experimental difficulties associated both with the measurements and the interpretation of the results have, so far, prevented an elucidation of the problem and, in fact, contradictory results persist, with some experiments obtaining a finite non-zero result \cite{LWM14, ZMD16} and others compatible with instantaneous tunneling \cite{TMK15}.  It should be noticed that the similarity between Schr\"odinger and Helmholtz equations allows for analogies between quantum tunneling of massive particles and photons \cite{CKS91}, and a non-instantaneous tunneling time is supported by this analogy and experiments measuring photonic tunneling times \cite{SKC93}, as well as by many theoretical calculations based on both the Schr\"odinger (for reviews see, e.g., \cite{Win06, Win06a}) and the Dirac equations (e.g., \cite{LSM98, KSG01, LCh02, PJa03, CLi03, LRo07, LMa07, LMN11b}).

The conceptual difficulty in obtaining an unambiguous and well-defined tunneling time is associated with the impossibility of obtaining a self-adjoint time operator in quantum mechanics \cite{Per80}, therefore leading to the need for operational definitions of time.  Several such definitions exist, such as phase time \cite{Wig55}, dwell time \cite{Smi60}, the Larmor times \cite{Baz67, Ryb67, But83, FHa88}, the Salecker-Wigner-Peres (SWP) time \cite{Per80, SWi58}, etc., and in some situations these lead to different, or even contradictory, results. This is not surprising, since by their own nature operational definitions can only describe limited aspects of the phenomena of tunneling, and it is unlikely that any one definition will be able to provide an unified description of the quantum tunneling times in a broad range of situations. Nevertheless, it remains an important task to obtain a well-defined and \emph{real} time scale that accurately describes the recent experiments \cite{EPC08, LWM14, OMP14, LKe15, POS15, TMK15, ZMD16}.

It is important to notice that the time-independent approach to tunneling times  (i.e., for incident particles with sharply defined energy), which comprises the vast majority of the literature, is ill-suited to accomplish the above mentioned goal, since it ignores the essential role of localizability in defining a time scale \cite{FHa88, LMN11a} -- see, however, \cite{KCN09}, which applies the time defined in \cite{GIA04} to investigate the half-life of $\alpha$-decaying nuclei.
A few works (e.g., \cite{FHa88, BSM94, POl06, LMN11a}) address the issue of localizability and, consequently, arrive at a probabilistic definition of tunneling times (that is, an average time). In particular, in \cite{LMN11a} the SWP clock was used to obtain an average tunneling time of transmission (reflection) for an incident wave packet, and such time was employed to investigate the Hartman effect \cite{Har62} for a particle scattered off a square barrier and it was shown that it does not saturates in the opaque regime \cite{LMN11a, FLM14}.

The tunneling time scales considered in \cite{LMN11a, FHa88, BSM94} involve taking an average over the spectral components of the transmitted wave packet and, thus, obscure the interpretation of the resulting average time. In this paper, we take as a starting point the real-valued average tunneling time obtained in \cite{LMN11a}, using the SWP quantum clock, and obtain a \emph{probability distribution of transmission times}, by using a standard transformation between random variables. In addition to providing a more accurate time characterization of the tunneling process, this should provide a clearer connection with the experiments (which measure a distribution of tunneling times -- see, e.g., Fig. 4 in \cite{LWM14}). It is worth noting that some approaches using Feynman's path integrals address the problem of obtaining a probabilistic distribution of the tunneling times (see, e..g., \cite{Tur14}). However, these methods in general result in a complex time (or, equivalently, multiple time scales), and some \emph{arbitrary} procedure is needed to select the physically meaningful real time \emph{a posteriori}.

After obtaining a general formula for the distribution of tunneling times, which is the main result of this work, we apply it to two specific cases. First, to illustrate the formalism in a simple scenario, we consider the situation of a particle tunneling through a rectangular barrier.  Then, we consider a slight modification of the model proposed in \cite{BSM10} for the tunneling decay of an initially bound state, after the sudden deformation of the binding potential by the application of a strong external field -- the modification considered here allows us to investigate the whole range of possibilities for the tunneling times, without having an ``upper cutoff",  as is the case in the original model. Finally, some additional comments on the results are reserved for the last section.



\section{The SWP clock's average tunneling time}

We start by briefly reviewing the time-dependent application of the SWP clock to the scattering of a massive particle off a localized static potential barrier in one dimension (for details see \cite{LMN11a}) -- which is appropriate, since it follows from the three-dimensional Schr\"odinger equation for this problem that the dynamics is essentially one-dimensional \cite{LWM14}.

The SWP clock is a quantum rotor weakly coupled to the tunneling particle and that runs only when the particle is within the region in which $V(x) \neq 0$, where $V(x)$ is the potential energy. The Hamiltonian of the particle-clock system is given by (we use $\hbar=2\mu =1$, where $\mu$ is the particle's mass) \cite{Per80}
\begin{equation}\label{Hc+s}
H=-\frac{\partial^2}{\partial x^2} + V(x)+ {\mathcal P}(x) H_c,
\end{equation}
where ${\mathcal P}(x)=1$ if $V(x)\neq0$ and zero otherwise. The clock's Hamiltonian is  $H_c=-i\omega \frac{\partial}{\partial\theta}$, where the angle $\theta\in [0,2\pi)$ is the clock's coordinate and $\omega=\frac{2\pi}{(2j+1)\vartheta}$ is the clock's angular frequency, with $j$ a non negative integer or half-integer giving the clock's total angular momentum and $\vartheta$ is the clock's resolution. The weak coupling condition amounts to assume that $\vartheta$ is large, in such a way that the clock's energy eigenvalues, $\eta_m \equiv m\omega$ ($-j<m<j$), are very small compared to the barrier height and the particle's energy. It is assumed that at $t=0$, well before it reaches the barrier, the particle is well-localized far to the left of the barrier and the wave function of the system is a product state of the form
\begin{equation}
\Phi(\theta,x,t=0)=\psi(x)v_0(\theta),
\end{equation}
where $\psi(x)$ is the particle's initial state, represented by a wave packet centered around an energy $E_0$, and the clock initial state is assumed to be ``in the zero-th hour" \cite{Per80}
\begin{equation}\label{inclock}
v_0(\theta)=\frac{1}{\sqrt{2j+1}}\sum_{m=-j}^{j}u_m(\theta),
\end{equation}
where $u_m(\theta)=\frac{\mathrm{e}^{im\theta}}{\sqrt{2\pi}}$ are the clock's eigenfunctions corresponding to the energy eigenvalues $\eta_m$.

The state $v_0(\theta)$ is strongly peaked at $\theta=0$, thus allowing the interpretation of the angle $\theta$ as the clock's hand, since for a freely running clock the peak evolves to $\omega t_c$, where $t_c$ is the time measured by the clock \cite{Per80}. Since here clock and particle are coupled according to (\ref{Hc+s}), when the particle passes through the region $V(x) \neq 0$ it becomes entangled with the clock, with the wave function for the entire system given by
\begin{eqnarray}\nonumber
\Phi(\theta,x,t)&\!=\!&\frac{1}{\sqrt{2j+1}}\sum_{m=-j}^{j}\Psi^{(m)}(x,t)u_m(\theta),\\
\label{tdsolgen}\\
\Psi^{(m)}(x,t)&\!=\!&\int_{0}^{\infty} dk  A(k)\psi^{(m)}_k(x)\mathrm{e}^{-iE t},
\nonumber
\end{eqnarray}
where $E$ is the incident particle's energy, $k=\sqrt{E}$, and $A(k)$ is the Fourier spectral decomposition of the initial wave packet $\psi(x)$ in terms of the free particle eigenfunctions (we are assuming delta-normalized eigenfunctions). The functions $\psi^{(m)}_k(x)$ satisfy a time-independent Schr\"odinger equation with a constant potential $\eta_m$ in the barrier region. Outside the potential barrier region and for a particle incident from the left, the (unnormalized) solution $\psi^{(m)}_k(x)$ of the time-independent Schr\"odinger equation is given by \cite{LMN11a}:
\begin{equation}\label{statsolgen}
\psi_k^{(m)}(x)=\left\{
\begin{array}{ll}
\mathrm{e}^{ikx}+ R^{(m)} (k)  \mathrm{e}^{-ikx},&x\leq-L\\
 T^{(m)} (k) \mathrm{e}^{ikx},&x\geq L,
\end{array}
\right.
\end{equation}
where $T^{(m)} (k)$ [$R^{(m)} (k)$] stands for the transmission (reflection) coefficient, and it is assumed, without loss of generality, that the potential is located in the region $-L<x<L$. Considering only the transmitted solution in (\ref{statsolgen}) and substituting it into the time dependent solution (\ref{tdsolgen}), it can be shown that for weak coupling
\begin{equation}\label{tdtrgen}
\Phi_{tr}(\theta,x,t)=\int_{0}^{\infty} dk\, A(k)T(k) \mathrm{e}^{i(kx-Et)}v_0\left(\theta-\omega  t_c^T (k) \right),
\end{equation}
where
\begin{equation}\label{stime}
 t_c^T (k)  =-\left(\frac{\partial \varphi_T^{(m)}}{\partial \eta_m}\right)_{\eta_m=0}
\end{equation}
is the stationary transmission clock time corresponding to the wave number component $k$ \cite{Per80, CLM09}. The transmission coefficient $T(k)$ corresponds to the stationary problem in the absence of the clock.

For \emph{tunneling} times one is interested only in the clock's reading for the \emph{post-selected} asymptotically transmitted wave packet. Thus, tracing out the particle's degrees of freedom, the expectation value of the clock's measurement can be defined, resulting in the average tunneling time \cite{LMN11a}
\begin{equation} \label{tav}
\left\langle  t_c^T   \right\rangle=\int dk\, \rho(k) t_c(k),\qquad \rho(k)=N\left|A(k)T(k)\right|^2,
\end{equation}
where $N=1/\int dk \, \left|A(k)T(k)\right|^2$ is a normalization constant and $\rho(k)$ is the probability density of finding the component $k$ in the transmitted wave packet. Similar expressions can be obtained for the reflection time.

\section{ The tunneling times distribution}

An important aspect of the average tunneling time considered in the previous section is that it emphasizes the \emph{probabilistic nature of the tunneling process}. However, since the average in (\ref{tav}) is over the time taken by the \emph{spectral components} of the wave packet, it does not lend itself to an easy interpretation, given that the spectral components of the wave packet tunnel with different times. Thus, instead of (\ref{tav}), one would rather obtain an average over (\emph{real}) times of the form
\begin{equation} \label{tavt}
\left\langle t_c \right\rangle=\int_0^{\infty} d\tau \, \tau\, \rho_t(\tau ) \; ,
\end{equation}
where $\rho_t(\tau)$ stands for the probability density for observing a particular tunneling time $\tau$ for the asymptotically transmitted wave packet.  This can easily be achieved by noticing that in probability theory (\ref{tav}) and (\ref{tavt}), which must be equal, are related by a standard transformation between the two random variables $k$ and $\tau$ through a function $t_c(k)$. It follows that the probability distribution of times is given by
\begin{equation}
\rho_t(\tau)= \int \rho(k)\delta\left(\tau-t_c^T(k) \right) dk
\label{rhot1}
\end{equation}
which, in essence, is the statement that all the $k$-components in the transmitted packet for which $t_c^T(k)=\tau$ must contribute to the value of $\rho_t(\tau)$ with a weight $\rho(k)$. Finally, using the properties of the Dirac delta function \footnote{Specifically, we use the fact that \\
$$\delta (g(x)) = \sum_j \frac{\delta(x-x_j)}{\left| g^\prime \left( x_j\right) \right| },$$ \\
where $\{ x_j \}$ is the set of zeros of the function $g(x)$ and the prime indicates a derivative with respect to the independent variable.}, we obtain
\begin{equation}
\rho_t(\tau)= \sum_{j} \frac{\rho\left(k_j(\tau)\right)}{\left| t_c^{T\prime}\left(k_j(\tau)\right) \right|},
\label{rhot2}
\end{equation}
where $\{k_j(\tau)\}$ is the set of zeros of the function $g(k)\equiv t^T_c(k)-\tau$ and $ t_c^{T\prime}$ is the derivative of $t^T_c(k)$ with respect to $k$.

A similar definition of the distribution of tunneling times given in (\ref{rhot1})-(\ref{rhot2}) can be obtained for any time scale which is probabilistic in nature, that is, of the form (\ref{tav}). Although several other probabilistic tunneling times exist in the literature (e.g., \cite{FHa88, BSM94, POl06, Tur14}), the SWP clock has proven to yield well-behaved \emph{real} times both in the time-independent \cite{Per80, CLM09, Par11} and time-dependent approaches \cite{Par09, LMN11a, FLM14} and it provides a simple procedure to \emph{derive} the probabilistic expression (\ref{tav}). In addition, the role exerted by circularly polarized light in attoclock experiments \cite{EPC08, LWM14}  seems to provide a natural possibility for interpretation in terms of the SWP clock.

As will be illustrated below, for the simple application of this formalism to the problem of a wave packet scattered off a rectangular potential barrier, the distribution of times (\ref{rhot1})-(\ref{rhot2}) cannot, in general, be obtained analytically even for the simplest cases -- except in trivial cases such as for a single Dirac delta potential barrier \cite{AER02, AER03, LLM16}, in which case $t_c^T(k) =0$ and $\rho_t(\tau) =\delta (\tau )\int dk\,\rho (k)$.

It should also be noticed that, despite the fact that the derivation of the previous section leading to (\ref{tav}) and, thus, (\ref{rhot1})-(\ref{rhot2}), assumed a scattering situation, these expressions can be shown to be valid for any situation involving pre-selection of an initial state localized to the left of a potential ``barrier" followed by post-selection of an asymptotic transmitted wave packet. This allows us to obtain the distribution of times for a model that simulates the tunneling decay of an initially bound particle by ionization induced by the sudden application of a strong external field -- the model considered below is a variant of that introduced in \cite{BSM10}.

\section{The distribution of tunneling times for a rectangular barrier}

\begin{figure*}
\includegraphics{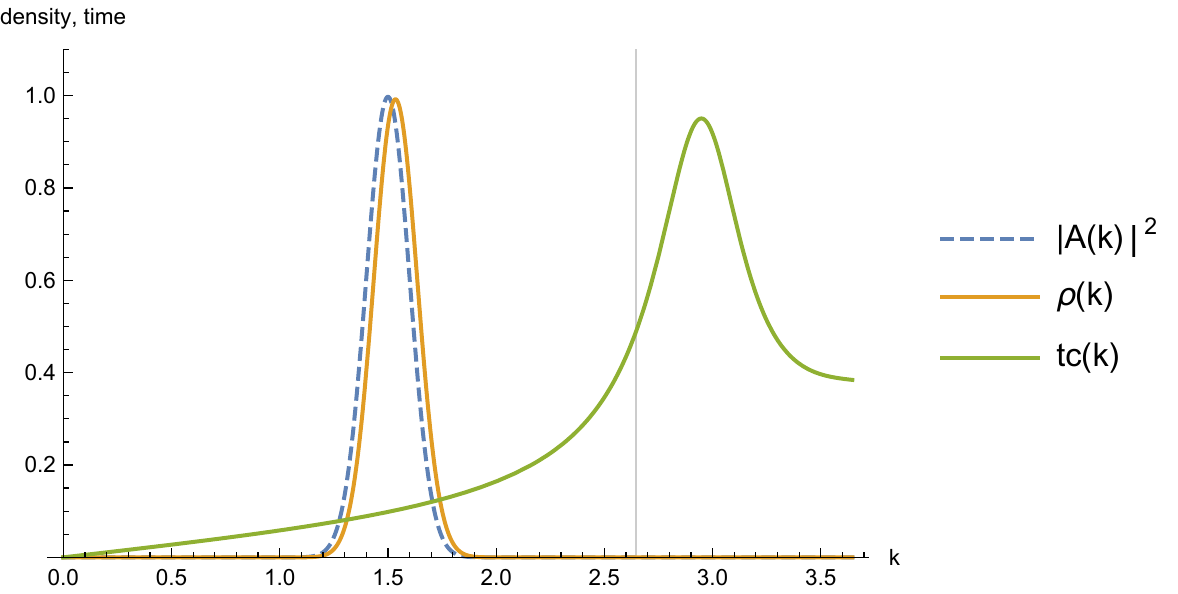}
\vspace{.5cm}
\includegraphics{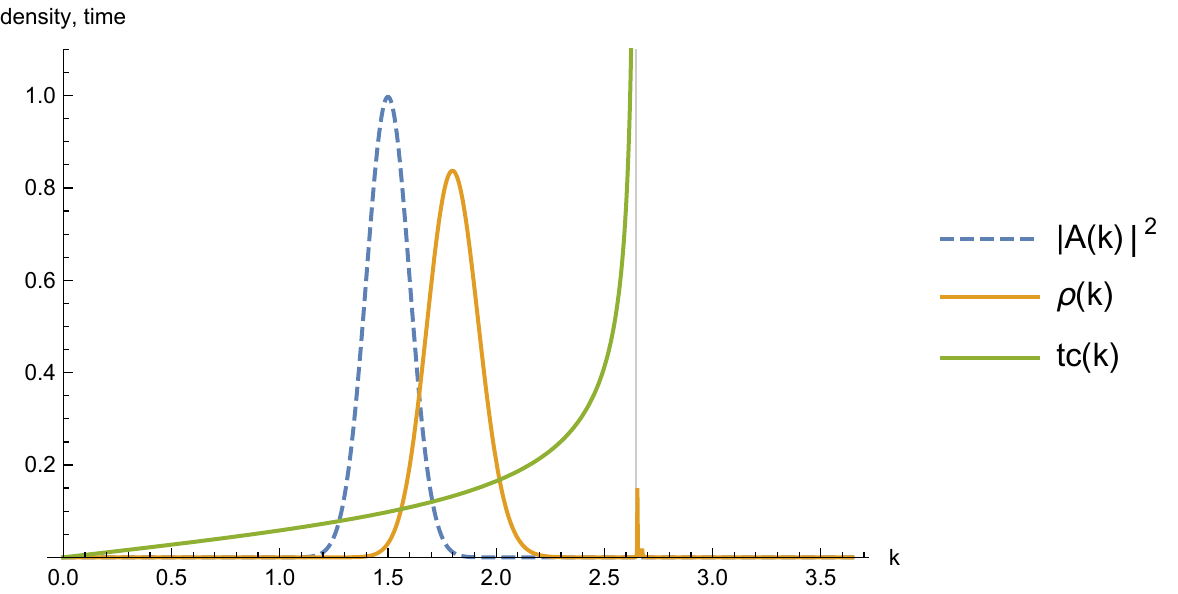}
\caption{\label{singfig1} Stationary transmission clock time $t_c(k)$ (green) and the distributions $\rho(k)$ (orange, arbitrary scale) and $\left|A(k)\right|^2$ (blue, dashed, arbitrary scale) for the transmitted and incident wave packets, respectively. Rydberg atomic units $\hbar=2m=1$ are used in all plots; the tunneling energies correspond to $0<k<\sqrt{7}$, corresponding to a barrier height $V_0=7$ (the maximum tunneling wave number $\sqrt{7}$ is shown by a vertical grey line in the plots). In both plots the incident wave packet parameters are $k_0=1.5, \sigma=5, x_0=-8\sigma$. \emph{Top:} barrier width $2L=2$. \emph{Bottom:} barrier width $2L=16$. }
\end{figure*}

As a first illustration of the formalism developed above, let us consider a rectangular barrier of height $V_0$ located in the region $x\in (-L,L)$. The particle's initial state $\phi_0(x) \equiv \psi (x, t = 0)$ is assumed to be a Gaussian wave packet
\begin{equation}
 \phi_0(x)  =\frac{1}{(2\pi)^\frac{1}{4}\sqrt{\sigma}} \exp\left[i k_0 x-\frac{(x-x_0)^2}{4\sigma^2} \right],
\end{equation}
where the parameters $x_0$, $\sigma$ and $k_0$ are chosen such that the wave packet is sharply peaked in a tunneling wave number $k_0=\sqrt{E_0}<\sqrt{V_0}$ and is initially well localized around $x=x_0$, far to the left of the barrier -- in the calculations that follow we take $x_0=-8\sigma$, such that at $t=0$ the probability of finding the particle within or to the right of the barrier is negligible. The transmission coefficient $T(k)$ and the spectral function $A(k)$ are well known, and given by
\begin{eqnarray}
T(k)&\!=\!&\frac{2 i k q e^{-2 i k L}}{\left(k^2-q^2\right) \sinh (2 L q)+2 i k q \cosh (2 L q)}\\
A(k)&\!=\!&\left(\frac{2}{\pi} \right)^\frac{1}{4} \sqrt{\sigma} \exp \left[4 k \sigma  \left(k_0 \sigma +4 i\right)\right. \nonumber \\
&&\left.-\sigma
   \left(k+k_0\right) \left(k \sigma +k_0 \sigma +8 i\right)\right],
\end{eqnarray}
where $q=\sqrt{V_0-k^2}$. The stationary transmission clock time (\ref{stime}) is  \cite{But83, LMN11a}
\begin{equation}\label{ct}
 t_c^T(k)  = \frac{k}{q} \frac{\left(q^2+k^2\right) \tanh(2 q L) + 2 q L \left( q^2 - k^2\right) \mathrm{sech}^2(2 q L)}{4 q^2 k^2+\left(q^2-k^2\right)^2 \tanh^2(2 q L)},
\end{equation}
with tunneling times corresponding to real values of $q$ (i.e., $V_0>k^2$). Figure \ref{singfig1} shows a plot for the stationary transmission times $t_c(k)$, the distribution of wave numbers $\rho(k)$ in the transmitted wave packet, and the distribution $\left|A(k)\right|^2$ of wave numbers (momenta) in the incident packet, for two values of the barrier width. For the chosen parameters and barrier widths both the incident and the transmitted wave packets have an energy distribution very strongly peaked in a tunneling component (in the bottom plot of Figure \ref{singfig1} the barrier is much more opaque than that in the top plot and we can observe that -- even if with a negligible probability for the parameters chosen for this plot -- in this situation some above-the-barrier components start to appear in the distribution of the transmitted wave packet. So, in order to consider mainly transmission by tunneling we must restrict the barrier widths to not too large ones). We also observe the very well-known fact that the transmitted wave packet ``speeds up" when compared to the incident particle \cite{LMN11a}. As a general rule, the larger is the barrier width (i.e., the more opaque is the barrier), the greater is the translation of the central component towards higher momenta. In which concerns the off resonance stationary transmission time, it initially grows with the barrier width, and saturates for very opaque barriers (the Hartman effect); on the other hand, it presents peaks at resonant wave numbers that grows and narrows with the barrier width -- for a detailed discussion see \cite{LMN11a}).

\begin{figure*}
\includegraphics{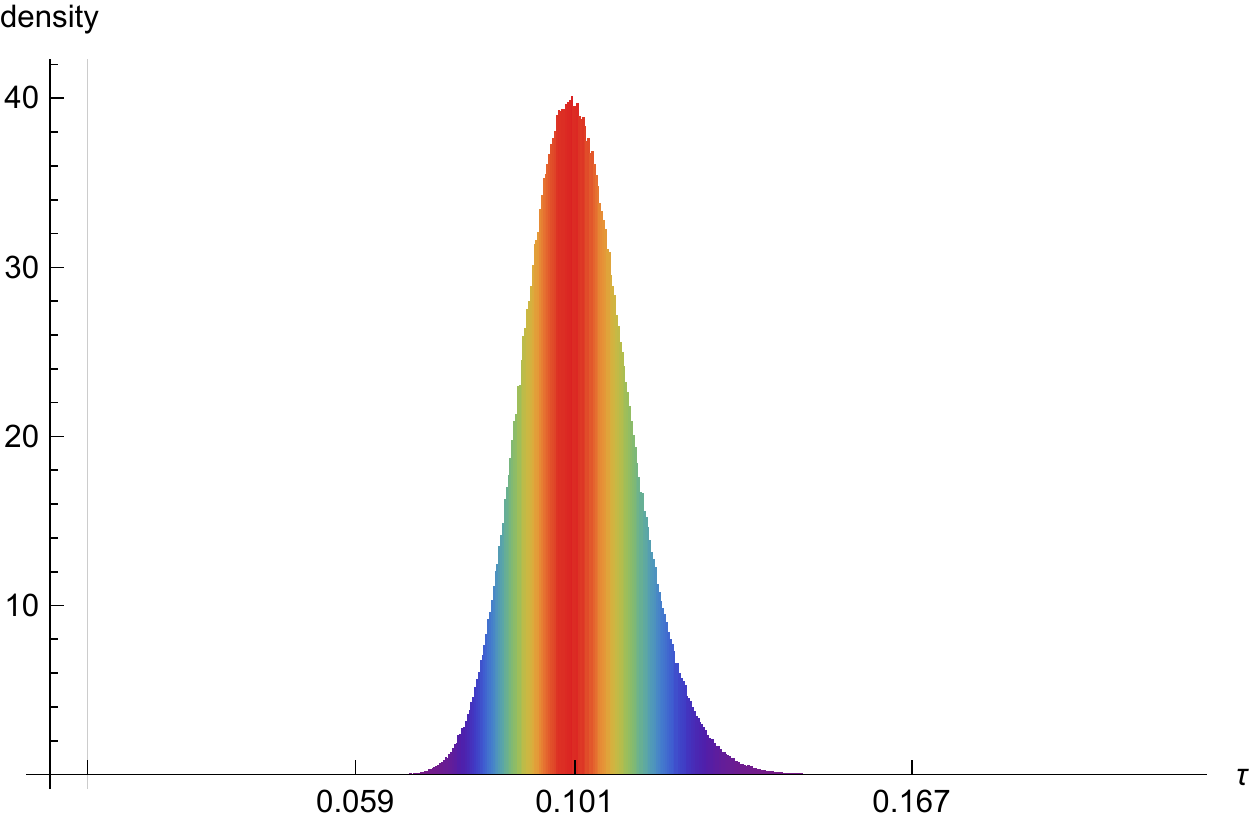}
\vspace{.5cm}
\includegraphics{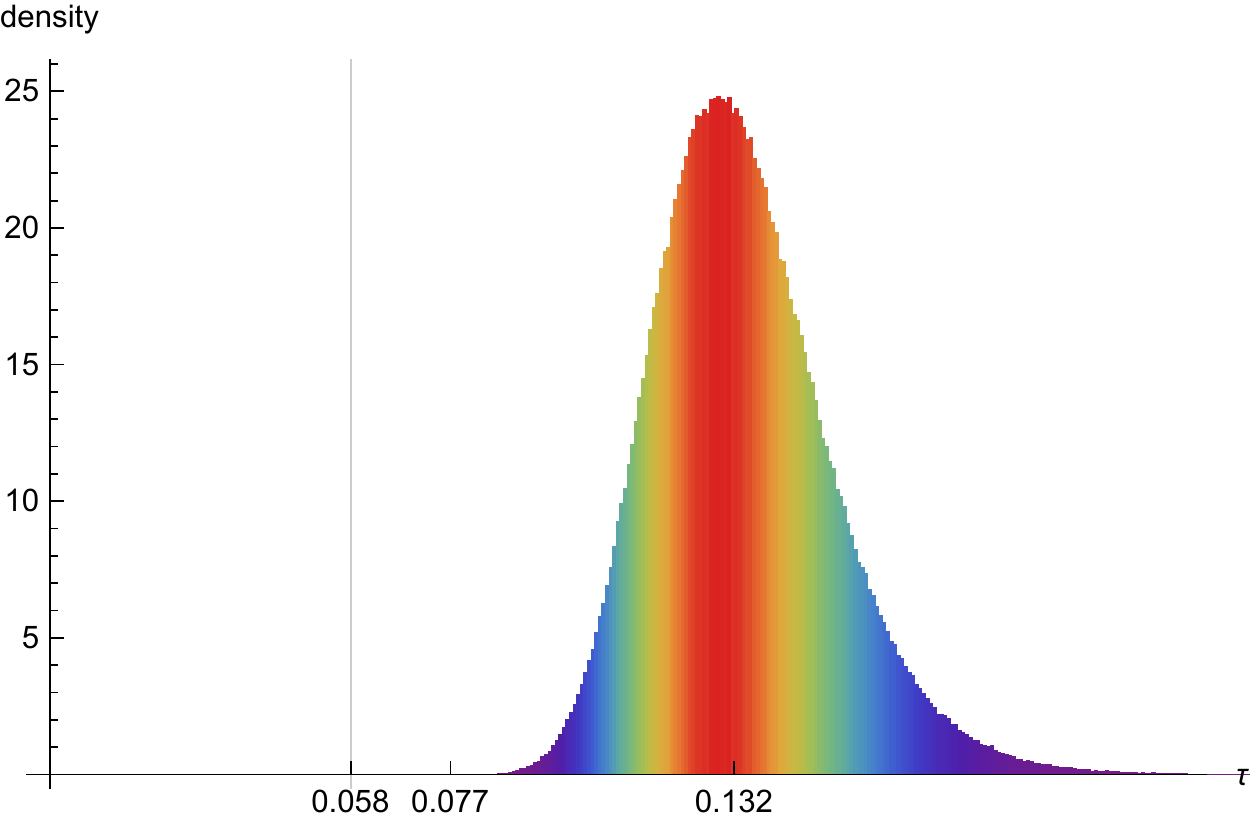}
\caption{\label{singfig2}  Probability distribution $\rho_t(\tau)$ for the tunneling times $\tau$, obtained by Monte Carlo samplings of $k$-values from the distribution $\rho(k)$ and then transforming these to time values through $\tau=t^T_c(k)$. The parameters are the same as in the corresponding plots in Figure \ref{singfig1}, and are all expressed in Rydberg atomic units. \emph{Top:} $2L=2$. \emph{Bottom:} $2L=16$. These plots are in the same range and scale and can be compared. The vertical grey line in both plots corresponds to the time the light takes to traverse the barrier distance. The ticks in the horizontal axes correspond to the light time, the minimum, the median and the maximum values of $\tau$ in the histogram (in the bottom plot the maximum $\tau$ is out of the plot's range).}
\end{figure*}

Figure \ref{singfig2} shows plots of the probability distribution $\rho_t(\tau)$ of the tunneling times according to Equations (\ref{rhot1})-(\ref{rhot2}), corresponding to both the barrier widths shown in the Figure \ref{singfig1} [to obtain these plots we used a Monte Carlo procedure to generate a large number of $k$ outcomes from the distribution $\rho(k)$, which afterwards were transformed into $\tau$ values by using the function $\tau=t_c(k)$].  The vertical grey lines in these plots correspond to the time the light takes to cross the barrier distance. It is observed that for the two distributions shown in Figure \ref{singfig2} the probability to observe superluminal tunneling times is negligible. It is also observed that these distributions have a shape that resembles that of the $k$ distribution, albeit with a more pronounced skewness. This shape could be inferred from Figure \ref{singfig1} and from Equation (\ref{rhot2}), since $t^{\prime}_c(k)$ grows very smoothly in the region were $\rho(k)$ is non-vanishing. Furthermore, a comparison between the two plots in Figure \ref{singfig2} shows that the tunneling times do not grow linearly with the barrier width and, therefore, the distribution in the bottom plot of Figure \ref{singfig2} is ``closer" to the light time than the distribution shown in the top plot; reference \cite {LMN11a}  already observed that for intermediate values of barrier widths the average transmission time -- corresponding to the mean of the distribution $\rho_t$ -- reaches a plateau.
%

\section{Distribution of ionization tunneling times}
\label{ionization}

In this Section we obtain a distribution for tunneling times for a particle that is initially in a bound state of a given binding potential. The potential is then suddenly deformed in such a way that the particle can escape from the initially confining region by tunneling. The model considered here is a slight modification of that proposed by Ban \textit{et al.} \cite{BSM10} to simulate, in a simple scenario, key features of the decay of a localized state by tunneling ionization induced by the application of a strong external field with a finite duration.

In \cite{BSM10}, for  $t<0$, the particle is in an eigenstate of a semi-infinite square-well potential $V_1(x)$,
\begin{equation}\label{V1}
V_1(x)=\left\{
\begin{array}{ll}
+ \infty & x< 0 \\
0& 0\leq x \leq a \\
V_0& x>a
\end{array}
\right.,
\end{equation}
\noindent
and, therefore, it cannot decay by tunneling. At $t=0$ the potential is suddenly deformed to $V_2(x)$,
\begin{equation}\label{V2}
V_2(x)=\left\{
\begin{array}{ll}
+ \infty & x< 0 \\
0& 0\leq x \leq a \\
V_0& a < x < b \\
0& x\geq b
\end{array}
\right.,
\end{equation}
such that the particle can now tunnel through the potential barrier -- it is assumed that the wave function does not change during the sudden change of the potential. Finally, after a finite time $t_0$ the potential returns to its original configuration, $V_1(x)$, and tunneling terminates. The cutoff time $t_0$ mimics the natural upper bound for tunneling times measured in recent attoclock experiments (see, e.g., \cite{LWM14, ZMD16} and references therein), since the opening and closing of the tunneling channel in these experiments occurs in intervals of half the laser field's period.

Here, we deviate from \cite{BSM10} by setting $t_0\to\infty$, i.e., once deformed the potential does not return to its original form and, after a long enough time, the particle will be transmitted with unit probability -- thus, by eliminating the cutoff (which is just an experimental limitation) we are able to explore the whole range of possibilities for the ionization tunneling time. In addition, for $t \geq 0$, the particle is assumed to be coupled to a SWP quantum clock running only in the region $(a,b)$, so that the clock's readings for the asymptotic transmitted wave packet give the time the particle spent within the barrier after $t=0$. Following \cite{BSM10}, we assume that for $t<0$ the particle is in the ground state of the potential $V_1(x)$, whose stationary wave function is given by
\begin{equation}\label{eqinstate}
\phi_0(x)=N \left\{\begin{array}{ll}
\sin \left(k_0 x\right),& 0<x\leq a\\
\sin k_0\; \mathrm{e}\,^{q_0(a-x)},&x>a
\end{array}\right. ,
\end{equation}
where $N$ is a normalization constant, $k_0=\sqrt{E_0}$, $E_0$ is the ground state energy, and $q_0=\sqrt{V_0-k_0^2}$. It is also assumed, as in \cite{BSM10}, that immediately after the sudden deformation of the potential from $V_1(x)$ to $V_2(x)$, at $t=0$, the wave function does not change. However, for $t\geq 0$ the particle state, which is no longer an energy eigenstate, is given by a superposition of the energy eigenstates $\psi_k(x)$  ($k=\sqrt{E}$) of the potential $V_2(x)$, i.e., \cite{BSM10}
\begin{equation}
\label{insuper}
\psi(x,t=0)=\phi_0(x)=\int_0^\infty S(k) \psi_k(x) \,dk,
\end{equation}
where
$$
S(k)=\int_0^\infty \phi_0(x) \psi_k^{*}(x)\, dx\, ,
$$
with
\begin{equation}\label{statv2}
\psi_k(x)=\left\{
\begin{array}{ll}
A(k)\sin(k x),& 0<x\leq a\\
C(k) \mathrm{e}\,^{q x}+D(k) \mathrm{e}\,^{-q x},& a<x\leq b\\
\sqrt{\frac{2}{\pi}} \cos [k (x - b) + \Omega(k)],& x>b
\end{array}
\right.,
\end{equation}
where $q=\sqrt{V_0-k^2}$ and the coefficients $A(k), C(k), D(k)$ and the phase $\Omega (k)$ are determined by the usual boundary conditions at $x=a$ and $x=b$, and are such that the normalization $\langle \psi_k(x),\psi_{k^\prime}(x)\rangle=\delta\left(k-k^{\prime}\right)$ holds \cite{BSM10}. From the above expressions it follows that, without any loss of generality, we can take $S(k)$ and all the eigenfunctions (\ref{statv2}) to be real.

In order to consider the coupling with the SWP clock for times $t\geq 0$ we proceed as follows. At $t=0$ the system  particle+clock is described by the product state $\psi(x,0)v_0(\theta)$, where $\psi(x,0)$ is the state (\ref{insuper}) and $v_0(\theta)$ is the initial clock state given by (\ref{inclock}). After $t=0$ the particle and the clock states become entangled. For the procedure of post selection of the asymptotically transmitted wave function we notice that the role of the transmission coefficient for the wave function (\ref{statv2}) is played by $\sqrt{2/ \pi} \; \mathrm{e}\,^{i\left(-kb+\Omega^{(m)}(k)\right)}$, where the superscript $m$ indicates the weak coupling with the clock. The right moving \emph{asymptotic} wave packet representing the coupled system formed by the transmitted particle and the clock is
\begin{eqnarray*}
\Phi_{tr}(\theta,x,t)&=& \int_0^\infty  dk\, S(k)  \, \mathrm{e}\,^{i [k (x-b) +\Omega^{(m)}(k) - E t]}\\
&&\times v_0[\theta-\omega t^T_c(k)]\, ,
\end{eqnarray*}
where, as before, $t^T_c(k)=-\left(\frac{\partial \Omega^{(m)}(k)}{\partial \eta_m}\right)_{\eta_m=0}=-\frac{1}{2 q}\frac{\partial \Omega}{\partial q}$ [with quantities without the subscript ``$(m)$" representing the limit $\eta_m \rightarrow 0$]. By following the same steps described in reference \cite{LMN11a}, we trace out the clock's degree of freedom in the asymptotic transmitted wave packet in order to obtain the distribution $\rho(k)$ of the wave numbers for the asymptotically transmitted wave packet, which in this case is simply given by
\begin{equation}\label{kdension}
\rho(k)=\left|S(k)\right|^2\, ,
\end{equation}
i.e., the probability to find a wave number $k$ in the asymptotic transmitted wave packet is the same as in the initial state, which is as expected, since after a long enough time the initial wave packet will be transmitted with probability unit, as mentioned earlier.

\begin{figure*}
\includegraphics{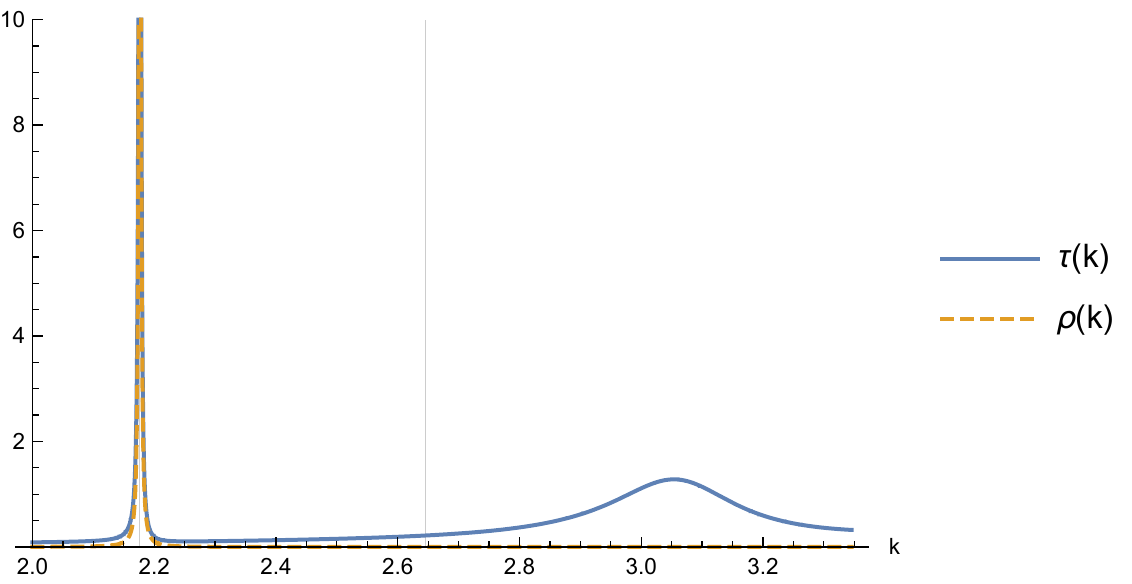}
\vspace{.5cm}
\includegraphics{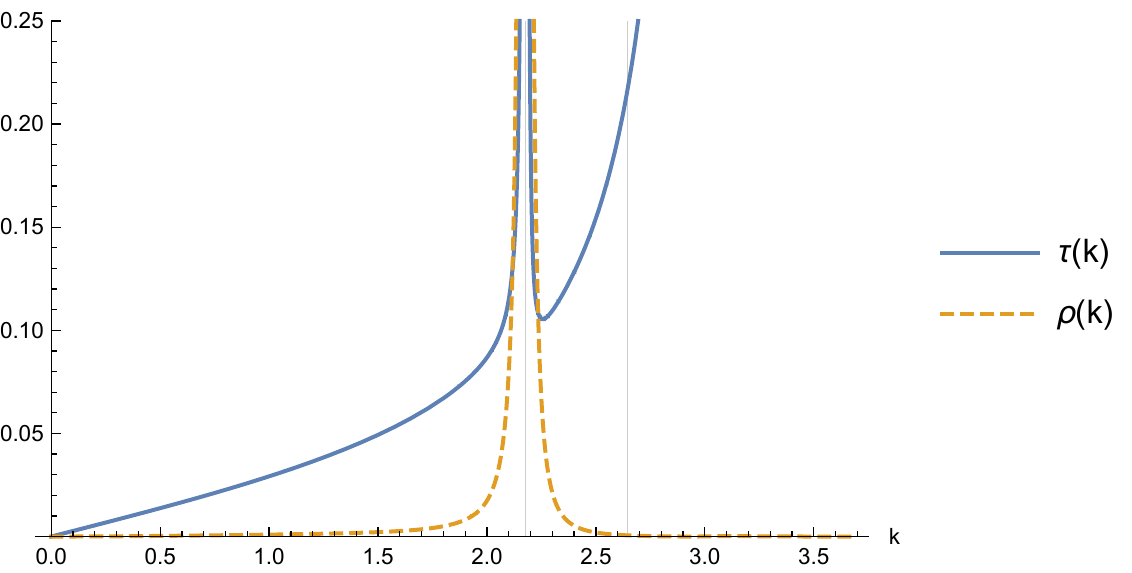}
\vspace{.5cm}
\caption{\label{spectral} \emph{Top}:  The stationary transmission clock time $t_c^T(k)$ (blue) and the wave number distribution $\rho (k)=\left|S(k)\right|^2$  (orange, dashed, arbitrary scale), for $V_0=7$, $a=1$, $b=3$ and $k_0\approx 2.175932$), with the initial state given by (\ref{eqinstate}). \emph{Bottom:} Close view of the above plot for small times. The vertical grey lines in the plots correspond to $k=k_0$ and $k=\sqrt{V_0}$. The regions in which $t_c^{T\prime}(k)\approx 0$ (around the local maximum and minimum of $t_c^T(k)$) correspond to times $\tau\approx t_c^T(k_0)$ and $\tau \approx 0.105\, a.u.$. Rydberg atomic units were used in all the plots.}
\end{figure*}

\begin{figure*}
\includegraphics{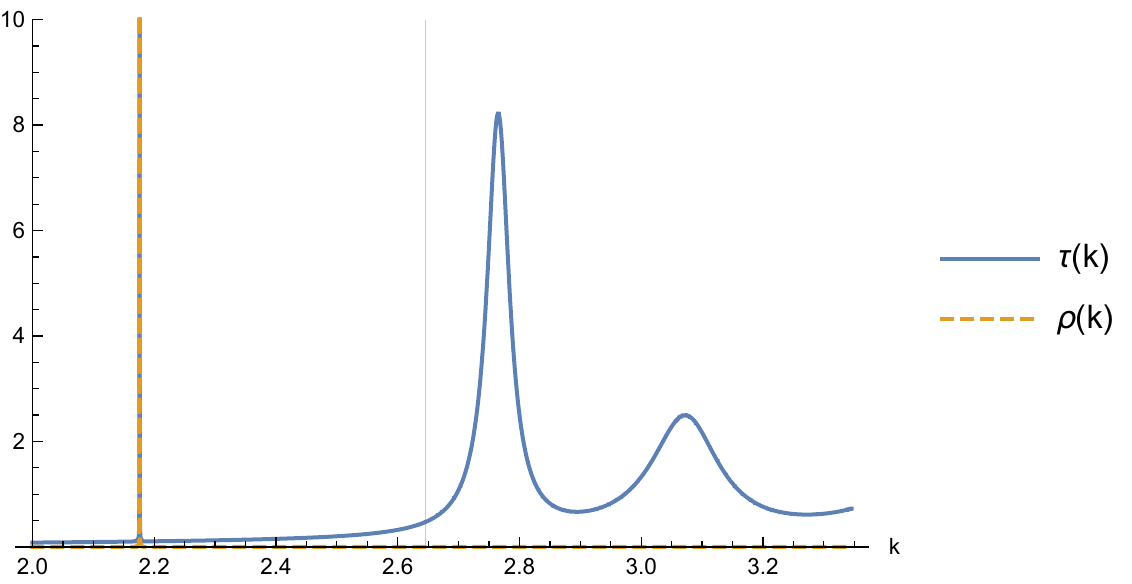}
\vspace{.5cm}
\includegraphics{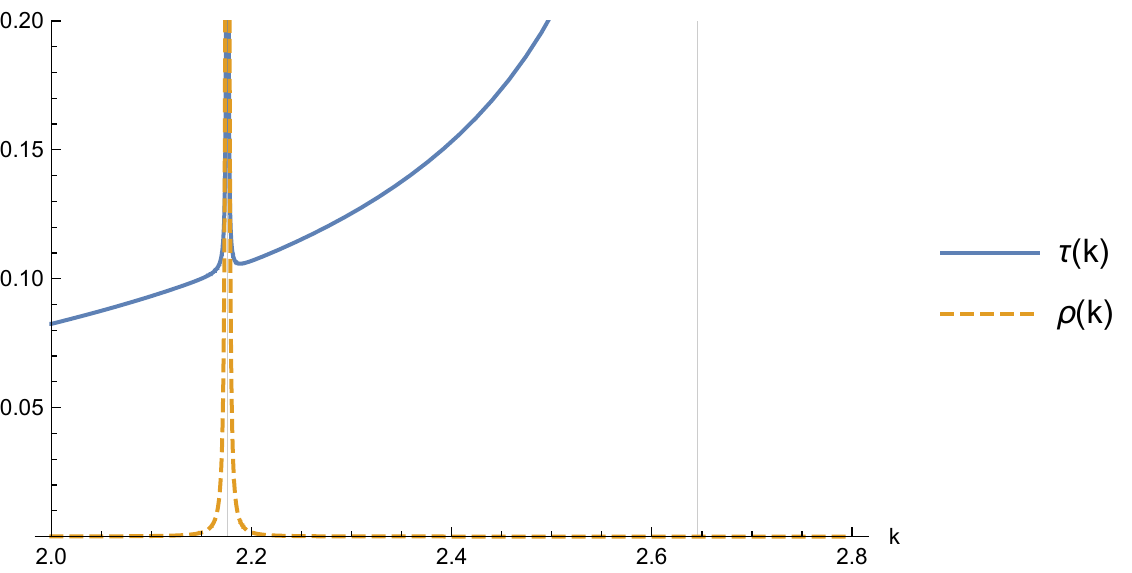}
\vspace{.5cm}
\caption{\label{spectral2} \emph{Top}:  The stationary transmission clock time $t_c^T(k)$ (blue) and the wave number distribution $\rho (k)=\left|S(k)\right|^2$  (orange, dashed, arbitrary scale), for $V_0=7$, $a=1$, $b=5$ and $k_0\approx 2.175932$), with the initial state given by (\ref{eqinstate}). \emph{Bottom:} Close view of the above plot for small times. The vertical grey lines in the plots correspond to $k=k_0$ and $k=\sqrt{V_0}$. The region of relatively slow growth of the derivative $t^{T\prime}(k)$ corresponds to times around $0.105\, a.u.$ Rydberg atomic units were used in all the plots.}
\end{figure*}

\begin{figure*}
\includegraphics{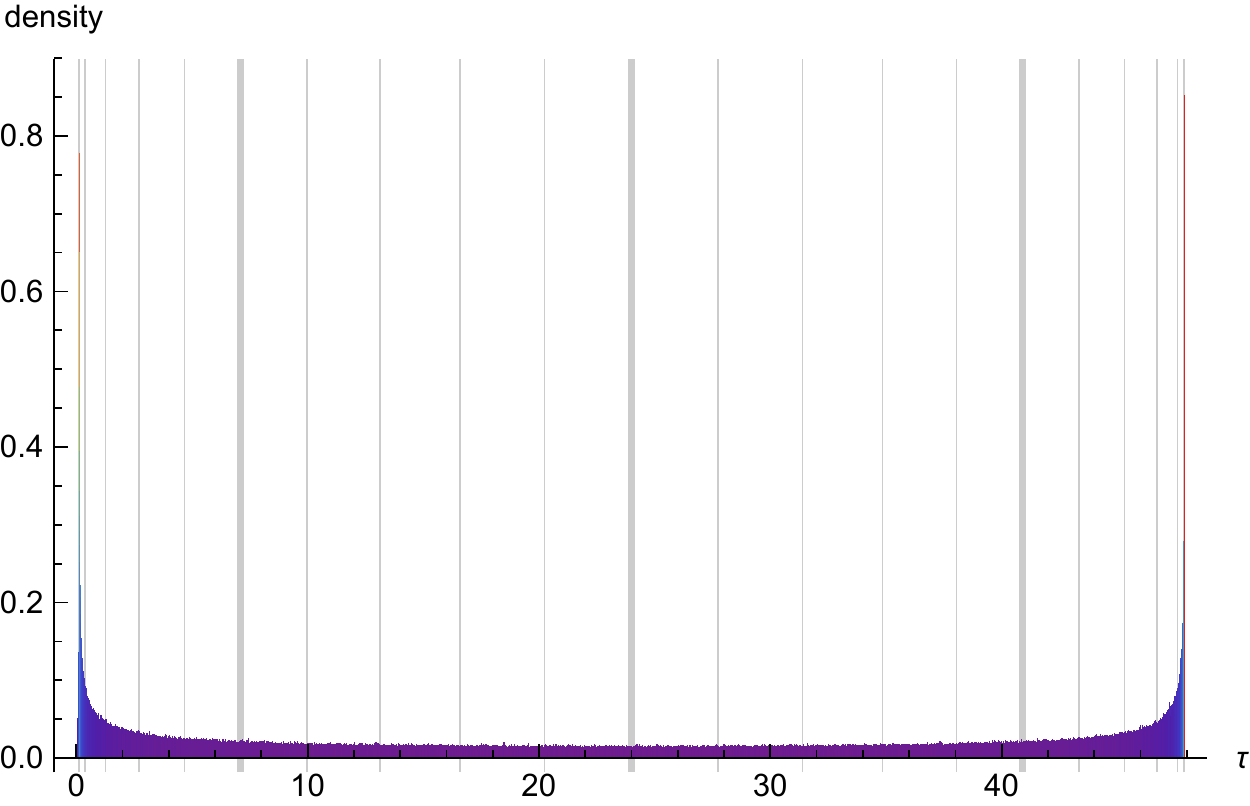}
\vspace{.5cm}
\includegraphics{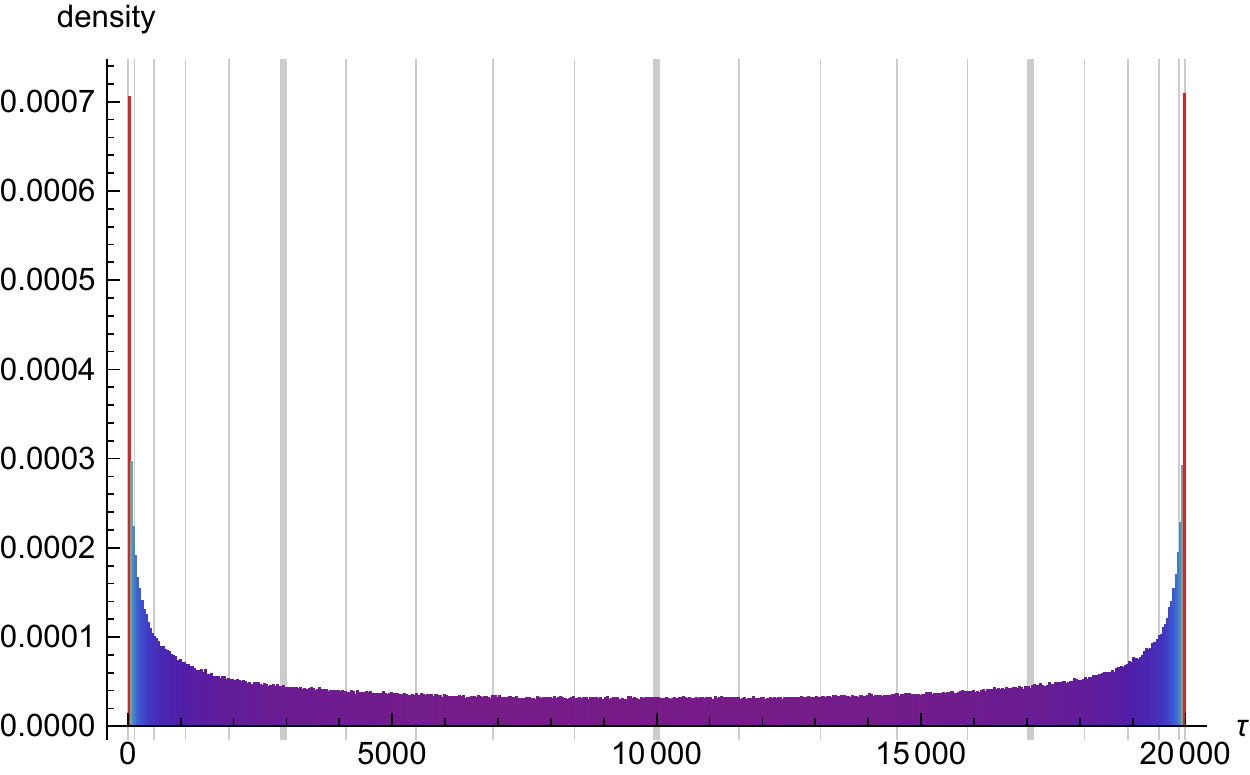}
\vspace{.5cm}
\caption{\label{histotais} Distributions of tunneling decay times $\rho_t(\tau)$ through the barrier of the potential $V_2(x)$ for the initial bound state $\phi_0(x)$ given by Equation (\ref{eqinstate}). The histograms were built by using the Monte Carlo procedure described in Figure \ref{singfig2} and in the main text. The vertical grey lines indicate percentiles of the distribution (the first and the last correspond to $1\%$ and $99\%$, the remaining ones range from $5\%$ to $95\%$, in steps of $5\%$); the three thick vertical lines indicate the first quartile (percentile $25\%$), the median and the third quartile (percentile $75\%$). Rydberg atomic units were used in all the plots. \emph{Top:} barrier width $b-a=2$, and bin length $\approx 0.0031 a.u.$ ($\approx 0.15$ attoseconds).  \emph{Bottom:} barrier width $b-a=4$, and bin length $\approx 40 a.u.$ ($\approx 1,935$ attoseconds).}
\end{figure*}

The general behavior of $t^T_c(k)$ and $\rho(k)$ is illustrated in Figures \ref{spectral} and \ref{spectral2}, corresponding to two barriers with different opacities ($b-a=2$ and $4$, respectively). These plots show, as expected, that the distribution $\rho(k)$ is strongly peaked at the wave number $k_0$, corresponding to the energy of the initially bound state, and is negligible for non-tunneling components. For tunneling wave numbers ($k<\sqrt{V_0}$) the function  $t_c^T(k)$ is also strongly peaked at the same wave number $k_0$, which corresponds to a local maximum (for non tunneling wave numbers there are several other resonance peaks). From Equation (\ref{rhot2}) we would expect that the peaks in the tunneling times distribution $\rho_t(\tau)$ would occur for times $\tau=t^T_c(k)$ corresponding to values of $k$ for which $t_c^{T\prime}(k)\approx 0$ -- which occur at points of local maxima and minima of the function $t^T_c(k)$ -- \emph{and} corresponding to non-negligible $\rho(k)$. Therefore, from the plots in Figures \ref{spectral} and \ref{spectral2} one could expect the first peak of the tunneling time distribution $\rho_t(\tau)$ at $\tau\approx 0.105\, a.u.$ (the local minimum of $t^T_c(k)$, which is similar for both barrier widths, since non-resonant times $t^T_c(k)$ change little with the barrier width for opaque barriers, as is the case in Figures \ref{spectral} and \ref{spectral2}); a second peak in $\rho_t(\tau)$ is expected to occur around the local maximum of $t^T_c(k)$, which corresponds to $\tau\approx t^T_c(k_0)$ (this local maximum -- corresponding to resonant wave numbers -- changes significantly with the barrier widths, see, e.g., \cite{LMN11a}). On the other hand, peaks in $\rho_t(\tau)$ coming from local maxima (resonances) and minima associated with non-tunneling values of $k$ are suppressed, since $\rho(k)\approx 0$ in these cases. Figure \ref{histotais} confirm these claims. For both barrier widths considered, the distribution of tunneling times is ``U" shaped, having peaks at the times corresponding to the local maxima and minima of the stationary time $t_c^T(k)$ inside the tunneling region. It should be observed that the larger is the barrier width, the broader is the tunneling time distribution,  due to the strong increase of the resonant tunneling time with the barrier width.

\begin{figure*}
\includegraphics{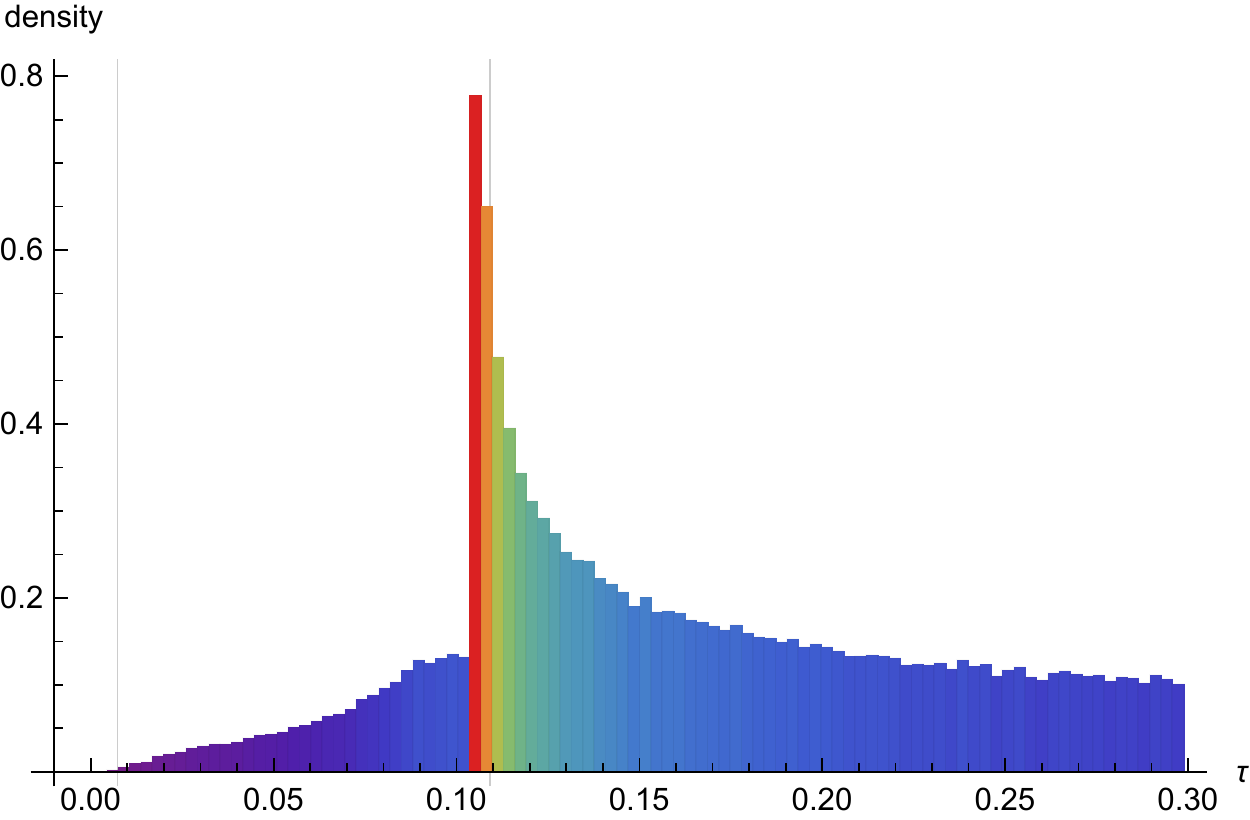}
\vspace{.5cm}
\includegraphics{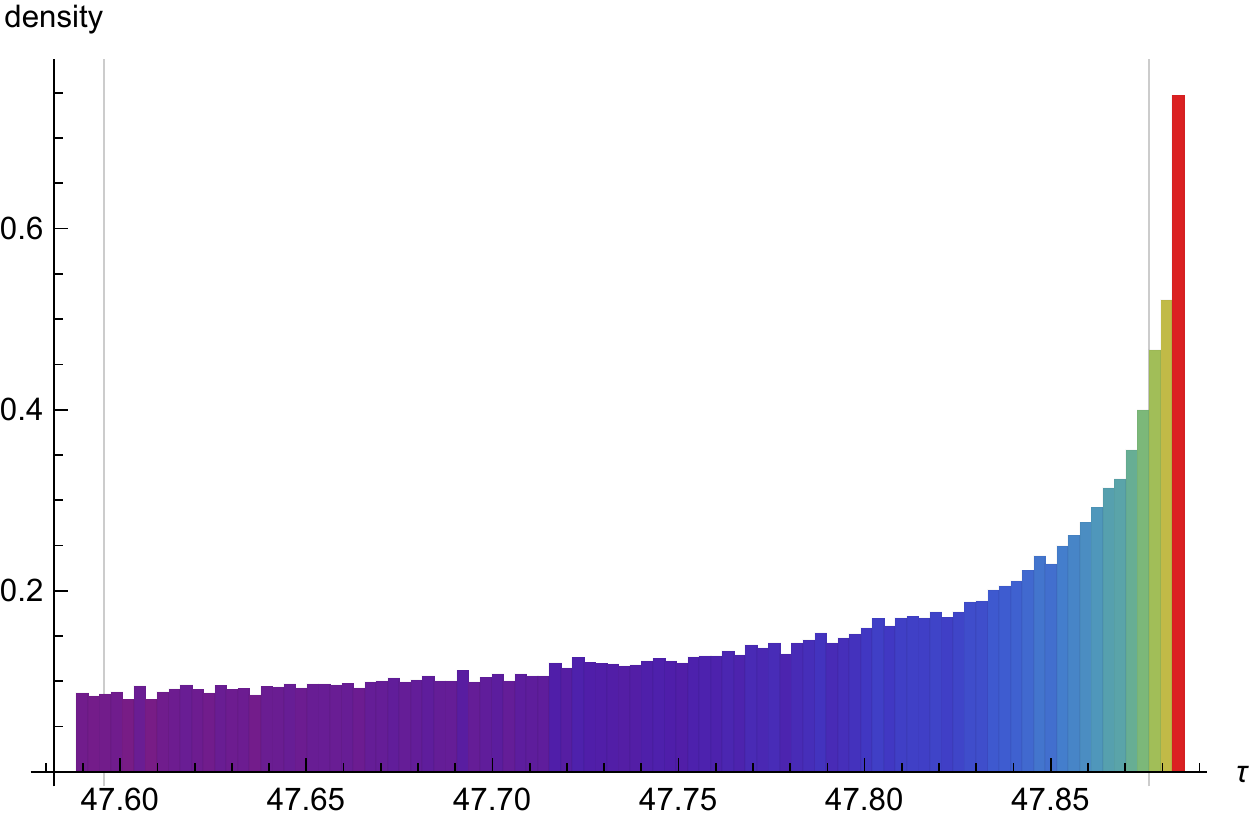}
\vspace{.5cm}
\caption{\label{close3} Close views of the plot at the top of Figure \ref{histotais}, corresponding to the barrier width $b-a=2$, with the bin length $\approx 0.0031 a.u.$ ($\approx 0.15$ attoseconds) . \emph{Top}: small tunneling times. The vertical grey line in the left of this plot corresponds to the time the light takes to travel the barrier distance. The second grey vertical line corresponds to the percentile $1\%$. \emph{Bottom:} large tunneling times. The vertical grey lines correspond to the percentiles $95\%$ and $99\%$, respectively.}
\end{figure*}

\begin{figure*}
\includegraphics{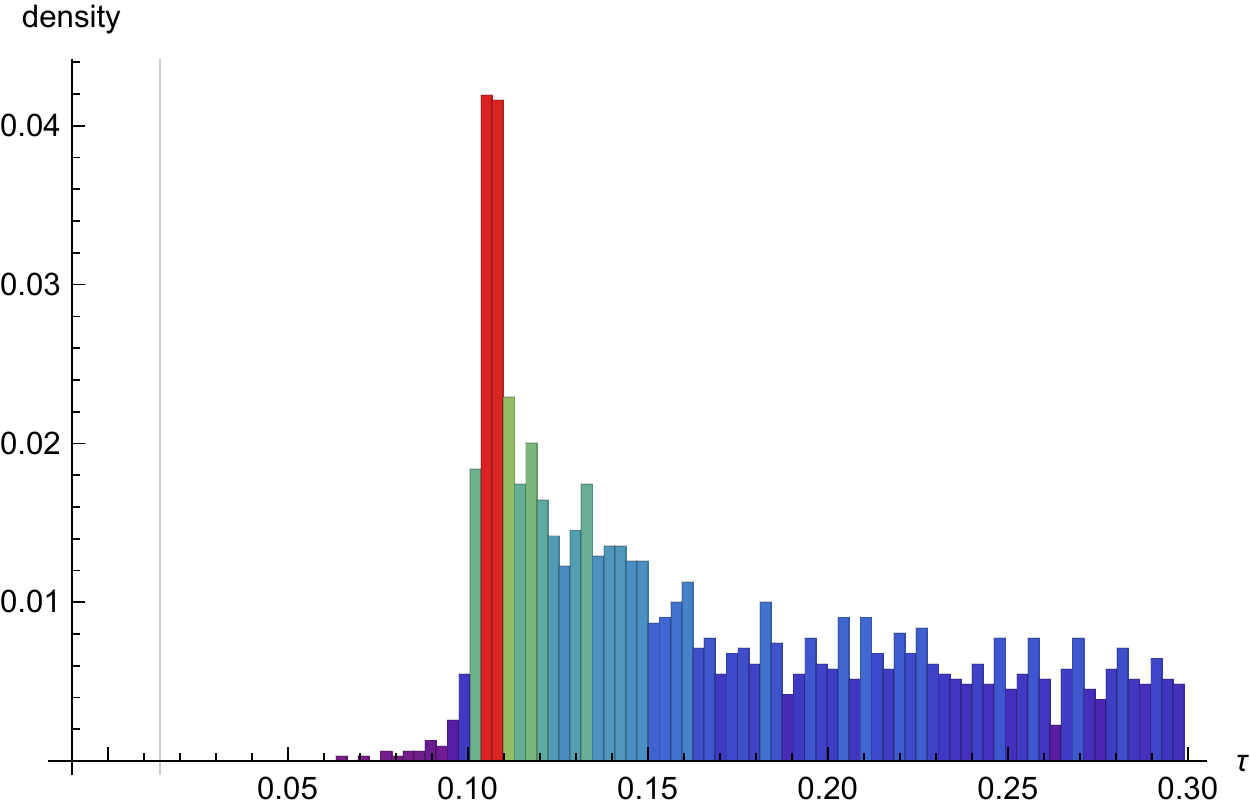}
\vspace{.5cm}
\includegraphics{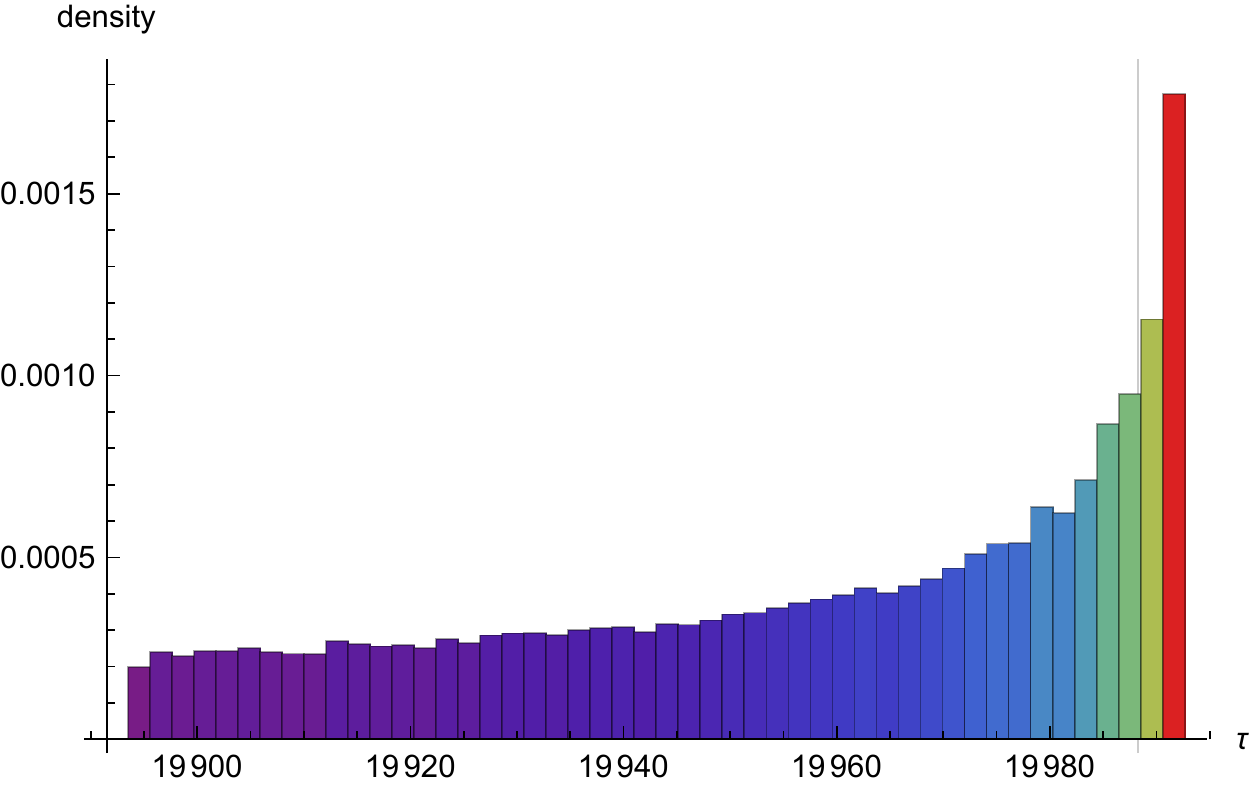}
\vspace{.5cm}
\caption{\label{close5} Close views of the plot at the bottom of Figure \ref{histotais}, corresponding to the barrier width $b-a=4$. \emph{Top}: small tunneling times and bin length $\approx 0.0031 a.u.$ ($\approx 0.15$ attoseconds). The vertical grey line in the left of this plot corresponds to the time the light takes to travel the barrier distance. The percentile $1\%$ (corresponding to $\approx 5.1\, a.u.\approx 247$ attoseconds) is out of the range of this plot. \emph{Bottom:} large tunneling times, with bin length $\approx 2\, a.u\approx 100$ attoseconds. The vertical grey line corresponds to the percentile $99\%$.}
\end{figure*}

Figures \ref{close3} and \ref{close5} show close views of the tunneling time distributions $\rho_t(\tau)$ for small and large tunneling times (Figure \ref{close3} corresponds to the plot at the top of Figure \ref{histotais}, whilst Figure \ref{close5} corresponds to the plot at the bottom of Figure \ref{histotais}). In the top plots of these Figures we can clearly observe the first peak around the local minimum of $t_c(k)$ in the tunneling region, which in both plots corresponds to almost the same value $\tau\approx 0.105\, a.u\approx 5.1$ attoseconds. The top plot of Figure \ref{close3} shows that for the less opaque barrier there exists a (very small) probability to observe a superluminal tunneling time. Even if this possibility cannot be precluded in principle, (see, e.g., \cite{LMN11b}), in the present case the possibility of emergence of such small times was expected, since at $t=0$ there was a significant portion of the wave packet (roughly $27\%$) penetrating the whole distance of the barrier, and this has an important contribution to the emergence of small times in the clock's readings associated to the transmitted particle. On the other hand, the top plot of Figure \ref{close5} shows that for the thicker barrier the probability for superluminal times is negligible -- the portion of the wave packet already inside the barrier at $t=0$ is the same ($\sim 27\%$), but the wave packet penetrates proportionally a smaller distance inside the barrier and, thus, it does not contribute in a significant way to the emergence of very small times in the clock readings. We note that the introduction of the cutoff $t_0$, as in \cite{BSM10}, would result in a time distribution similar to the truncated distributions shown in the top plots of Figures \ref{close3} and \ref{close5}.

\begin{figure*}
\includegraphics{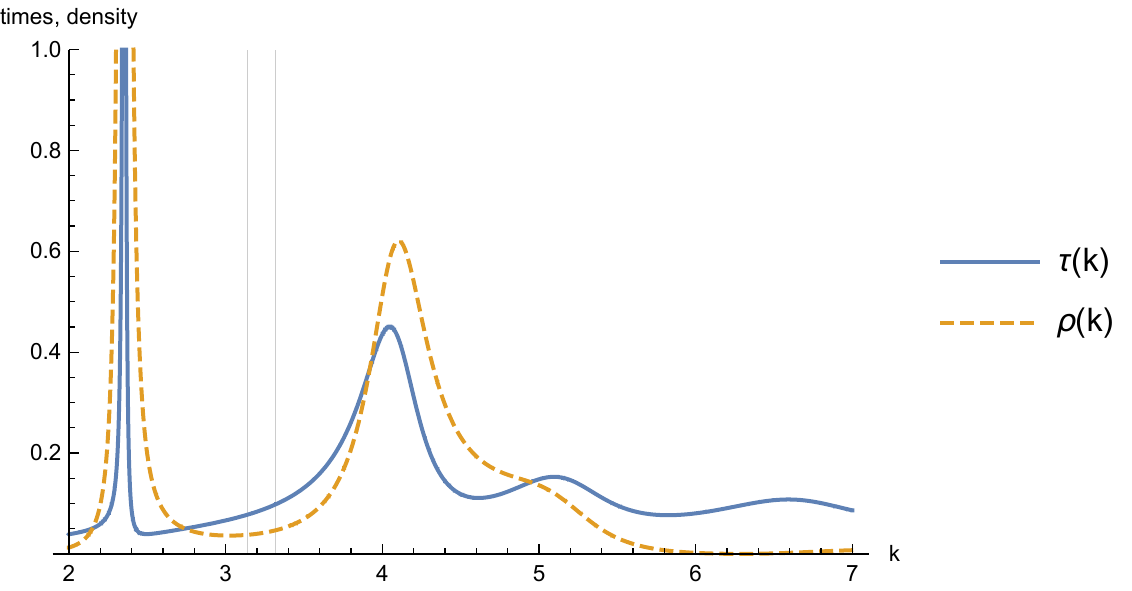}
\vspace{.5cm}
\includegraphics[scale=.95]{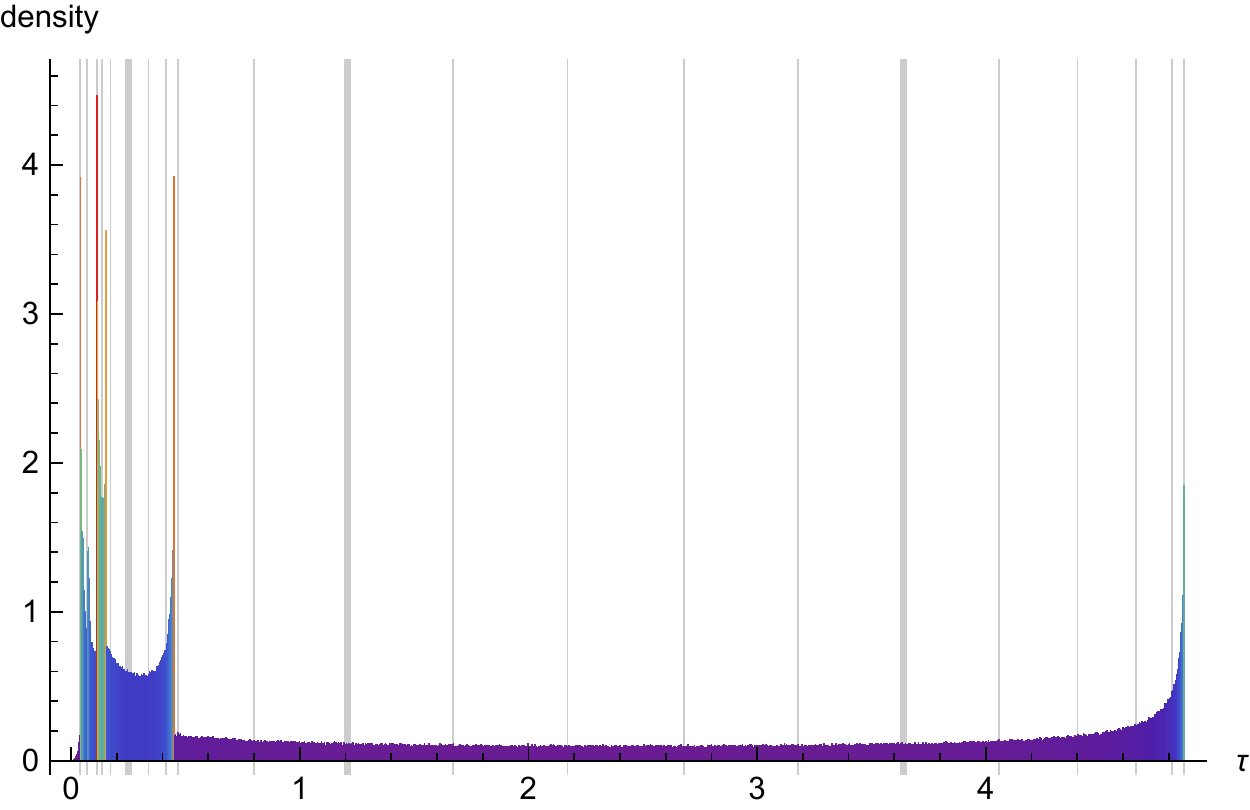}
\vspace{.5cm}
\includegraphics{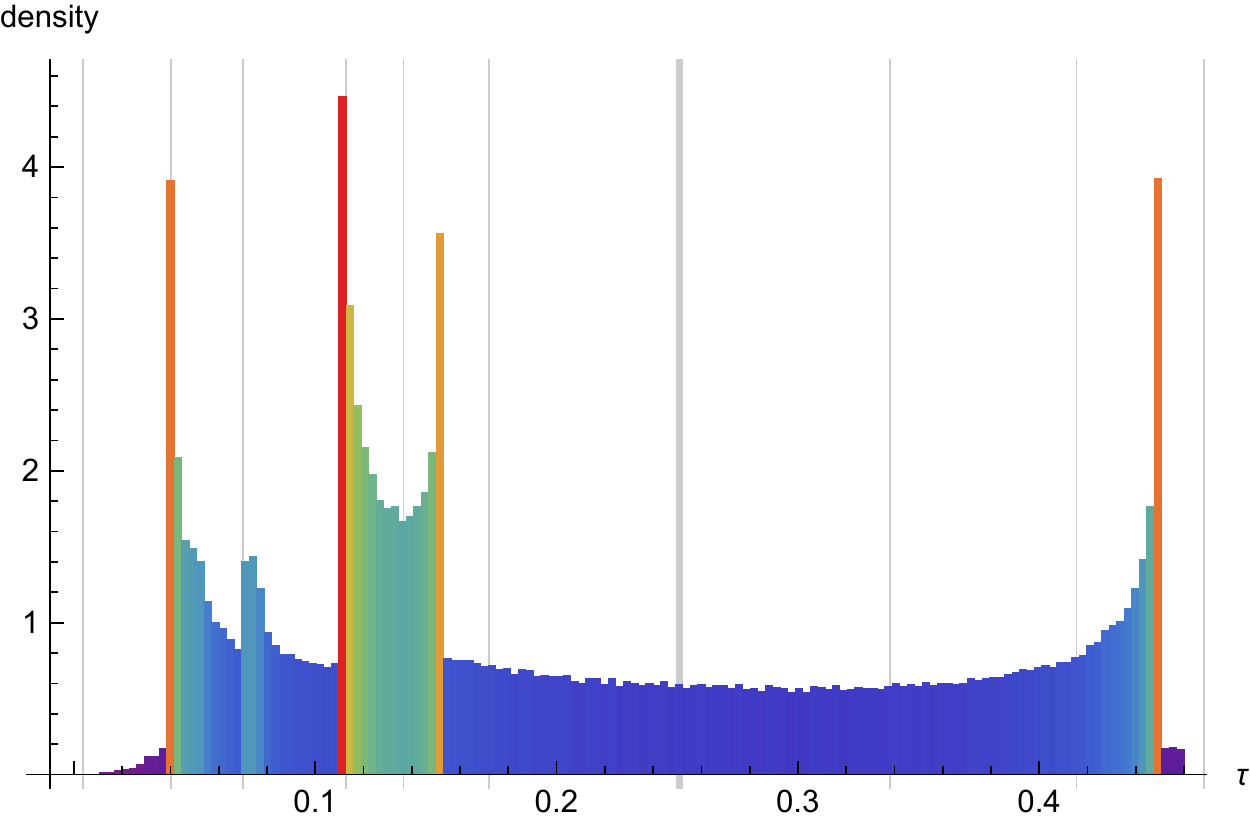}
\caption{\label{infig} \emph{Top}: The stationary time $t_c^T(k)$ and the wave number distribution $\rho(k)$, for $V_0=11$, barrier width $b-a=36$, $k_0=\pi$, and an initial state $\psi(x,0)=\sqrt{2}\sin k_0 x$, with $k_0=\pi$. The vertical lines correspond to $k=k_0$ and $k=\sqrt{V_0}$. \emph{Middle:} Distribution $\rho_t(\tau)$ for the transmission times, with the histogram built by the Monte Carlo procedure described in the text. Vertical lines indicate the percentiles, as in the Figure \ref{histotais}. \emph{Bottom:} Close view of the above histogram for the range of small times. The first vertical line at the left indicates the time the light takes to cross the barrier distance. In both the histograms we used a bin length $\approx 0.0031\, a.u.\approx 0.15$ attoseconds. Rydberg atomic units were used in all these plots.}
\end{figure*}

It is also worth to observe that, for small times, the distributions obtained here resemble qualitatively those in Figure 4 of \cite{LWM14}, except for the presence of several peaks at discrete values of the time in the latter. The considerations above, relating the peaks of the distribution of clock times $\rho_t(\tau)$ to the local maxima and minima of the stationary time $t_c^T(k)$ and the magnitude of distribution $\rho(k)$ in the neighborhood of these points, suggest a scenario in which such multiple peaks at discrete values of time can appear in the distribution $\rho_t(\tau)$ of transmission times. Indeed, if above-the-barrier wave numbers had a significant contribution to the initial wave packet, then the several local maxima and minima present in the vicinities of the resonant \emph{non tunneling} components will also contribute in a significant way to build multiple peaks in the distribution of transmission times; these peaks, however, could not be associated with the tunneling process. We can consider such a scenario by choosing as the initial state a tightly localized state given by $\psi(x,0)=\sqrt{2}\sin k_0 x$, with $k_0=\pi$ and the barrier parameters $a=1$, $b=2$ and $V_0=11$, in Rydberg atomic units. In this situation the initial wave function is perfectly confined to the left of the barrier ($0<x<1$), and above-the-barrier components contribute in a significant way to build the wave packet, as can be seen from $\rho(k)$ in the top plot of Figure \ref{infig} (in this case the probability of finding a non tunneling $k$ component in the wave packet is approximately $75\%$). In this plot we can also observe that all the local maxima and minima of $t^T_c(k)$ shown occur in neighborhoods of wave numbers $k$ for which $\rho(k)$ is non-negligible; therefore, all these local maxima and minima contribute significantly to build multiple peaks in the distribution of transmission times $\rho_t(\tau)$. The middle and the bottom plots of Figure \ref{infig} confirm this statement: all the peaks of the  distribution of transmission times correspond very closely to the local maxima and minima of $t_c(k)$, as can be seen by comparing the plots in the top and the bottom of this Figure (except for the first, all the other significant peaks in the bottom plot are associated to non-tunneling components).

%



\section{Conclusions}

Taking as a starting point the \emph{probabilistic} (average) tunneling time obtained in \cite{LMN11a} with the use of a SWP clock \cite{SWi58, Per80, CLM09}, we obtained a \emph{probability distribution of times}, equations (\ref{rhot1})-(\ref{rhot2}). An important advantage of using the SWP clock, in addition to those already mentioned, is that by running only when the particle is inside the barrier it allows us to address the concept of \emph{tunneling} time in a proper way, since the time spent by the particle standing in the well \emph{before} penetrating the barrier is \emph{not} computed. A clear advantage of having a probability distribution of transmission (tunneling) times is that in addition to the usual expectation value, we can obtain \emph{all} the statistical properties of this time, such as its most probable values (peaks of the distribution), the dispersion around the mean value, and the probability to observe extreme outcomes (superluminal times, for instance).

As an initial test, the distribution of times (\ref{rhot1})-(\ref{rhot2}) was applied to the simple problem of a particle tunneling through a rectangular barrier. Unsurprisingly, it revealed behavior similar to that already known from previous works using a distribution of wave numbers (momentum) -- see, e.g., \cite{LMN11a} -- although, using $\rho_t(\tau)$ these conclusions are much more transparent. For example, one could answer the question about the possibility of superluminal tunneling by direct calculation from the probability distribution $\rho_t(\tau)$. In the non stationary case -- which is the correct to address this question -- this problem is usually answered by considering just the average tunneling time. But,  given its probabilistic nature, an answer based only on the average time may not be satisfactory, especially if the dispersion of the distribution of tunneling times is large, which is often the case when one deals with well-localized particles, as suggested by the two situations addressed in the present work.

As a main application of (\ref{rhot1})-(\ref{rhot2}), we considered a slight modification of the problem considered in \cite{BSM10} to model strong field ionization by tunneling. The modification considered here was the elimination of the cutoff time that was introduced in \cite{BSM10} to simulate the upper bound that arises in attoclock experiments \cite{LWM14, ZMD16} due to the opening and closing of the tunneling channel, naturally associated with the oscillations in the laser field intensity. This cutoff is not a fundamental requirement, but rather it is associated to the experimental methods employed -- in any case, its implementation is rather trivial, since it just truncates the distribution of times. The consideration of the full range of the distribution of times allowed us to show that an important contribution to $\rho_t(\tau )$ comes from very large times associated with the resonance peaks in the tunneling region -- these very long tunneling times occur with a probability comparable to very short ones, thus having an important impact on the average tunneling times and, therefore, cause difficulties when comparing theoretical predictions based on an average time with the outcomes of experiments presenting a natural cutoff in the possible time measurements. In particular, in the attoclock experiments the relevant measure is often associated to the peak of the tunneling time, which may be promptly identified once one knows the probability distribution for all possible times. A remark is in place, the distribution of times proposed here, built on the SWP clock readings, refers to the time the particle \emph{dwells within} the barrier, while the tunneling times often measured in recent attoclock experiments actually refers to \emph{exit} times \cite{TYB16}.

In sum, the approach introduced above and resulting in (\ref{rhot1})-(\ref{rhot2}) builds upon the already (conceptually) well tested SWP clock to provide a real-valued distribution of times that, in the simple models considered here, was demonstrated to have physically sound properties and, in fact, (rough) similarities with the time distribution obtained in recent experiments \cite{LWM14}, therefore warrantying further investigation with more realistic potentials.



\bibliography{TunTimeDist[submitted]}

\begin{thebibliography}{44}%
\makeatletter
\providecommand \@ifxundefined [1]{%
 \@ifx{#1\undefined}
}%
\providecommand \@ifnum [1]{%
 \ifnum #1\expandafter \@firstoftwo
 \else \expandafter \@secondoftwo
 \fi
}%
\providecommand \@ifx [1]{%
 \ifx #1\expandafter \@firstoftwo
 \else \expandafter \@secondoftwo
 \fi
}%
\providecommand \natexlab [1]{#1}%
\providecommand \enquote  [1]{``#1''}%
\providecommand \bibnamefont  [1]{#1}%
\providecommand \bibfnamefont [1]{#1}%
\providecommand \citenamefont [1]{#1}%
\providecommand \href@noop [0]{\@secondoftwo}%
\providecommand \href [0]{\begingroup \@sanitize@url \@href}%
\providecommand \@href[1]{\@@startlink{#1}\@@href}%
\providecommand \@@href[1]{\endgroup#1\@@endlink}%
\providecommand \@sanitize@url [0]{\catcode `\\12\catcode `\$12\catcode
  `\&12\catcode `\#12\catcode `\^12\catcode `\_12\catcode `\%12\relax}%
\providecommand \@@startlink[1]{}%
\providecommand \@@endlink[0]{}%
\providecommand \url  [0]{\begingroup\@sanitize@url \@url }%
\providecommand \@url [1]{\endgroup\@href {#1}{\urlprefix }}%
\providecommand \urlprefix  [0]{URL }%
\providecommand \Eprint [0]{\href }%
\providecommand \doibase [0]{http://dx.doi.org/}%
\providecommand \selectlanguage [0]{\@gobble}%
\providecommand \bibinfo  [0]{\@secondoftwo}%
\providecommand \bibfield  [0]{\@secondoftwo}%
\providecommand \translation [1]{[#1]}%
\providecommand \BibitemOpen [0]{}%
\providecommand \bibitemStop [0]{}%
\providecommand \bibitemNoStop [0]{.\EOS\space}%
\providecommand \EOS [0]{\spacefactor3000\relax}%
\providecommand \BibitemShut  [1]{\csname bibitem#1\endcsname}%
\let\auto@bib@innerbib\@empty
\bibitem [{\citenamefont {Winful}(2006{\natexlab{a}})}]{Win06}%
  \BibitemOpen
  \bibfield  {author} {\bibinfo {author} {\bibfnamefont {H.~G.}\ \bibnamefont
  {Winful}},\ }\href {\doibase http://dx.doi.org/10.1016/j.physrep.2006.09.002}
  {\bibfield  {journal} {\bibinfo  {journal} {Physics Reports}\ }\textbf
  {\bibinfo {volume} {436}},\ \bibinfo {pages} {1} (\bibinfo {year}
  {2006}{\natexlab{a}})}\BibitemShut {NoStop}%
\bibitem [{\citenamefont {Winful}(2006{\natexlab{b}})}]{Win06a}%
  \BibitemOpen
  \bibfield  {author} {\bibinfo {author} {\bibfnamefont {H.~G.}\ \bibnamefont
  {Winful}},\ }\href {\doibase http://dx.doi.org/10.1088/1367-2630/8/6/101}
  {\bibfield  {journal} {\bibinfo  {journal} {New Journal of Physics}\ }\textbf
  {\bibinfo {volume} {8}},\ \bibinfo {pages} {101} (\bibinfo {year}
  {2006}{\natexlab{b}})}\BibitemShut {NoStop}%
\bibitem [{\citenamefont {Landsman}\ \emph {et~al.}(2014)\citenamefont
  {Landsman}, \citenamefont {Weger}, \citenamefont {Maurer}, \citenamefont
  {Boge}, \citenamefont {Ludwig}, \citenamefont {Heuser}, \citenamefont
  {Cirelli}, \citenamefont {Gallmann},\ and\ \citenamefont {Keller}}]{LWM14}%
  \BibitemOpen
  \bibfield  {author} {\bibinfo {author} {\bibfnamefont {A.~S.}\ \bibnamefont
  {Landsman}}, \bibinfo {author} {\bibfnamefont {M.}~\bibnamefont {Weger}},
  \bibinfo {author} {\bibfnamefont {J.}~\bibnamefont {Maurer}}, \bibinfo
  {author} {\bibfnamefont {R.}~\bibnamefont {Boge}}, \bibinfo {author}
  {\bibfnamefont {A.}~\bibnamefont {Ludwig}}, \bibinfo {author} {\bibfnamefont
  {S.}~\bibnamefont {Heuser}}, \bibinfo {author} {\bibfnamefont
  {C.}~\bibnamefont {Cirelli}}, \bibinfo {author} {\bibfnamefont
  {L.}~\bibnamefont {Gallmann}}, \ and\ \bibinfo {author} {\bibfnamefont
  {U.}~\bibnamefont {Keller}},\ }\href {\doibase
  http://dx.doi.org/10.1364/OPTICA.1.000343} {\bibfield  {journal} {\bibinfo
  {journal} {Optica}\ }\textbf {\bibinfo {volume} {1}},\ \bibinfo {pages} {343}
  (\bibinfo {year} {2014})}\BibitemShut {NoStop}%
\bibitem [{\citenamefont {Torlina}\ \emph {et~al.}(2015)\citenamefont
  {Torlina}, \citenamefont {Morales}, \citenamefont {Kaushal}, \citenamefont
  {Ivanov}, \citenamefont {Kheifets}, \citenamefont {Zielinski}, \citenamefont
  {Scrinzi}, \citenamefont {Muller}, \citenamefont {Sukiasyan}, \citenamefont
  {Ivanov},\ and\ \citenamefont {Smirnova}}]{TMK15}%
  \BibitemOpen
  \bibfield  {author} {\bibinfo {author} {\bibfnamefont {L.}~\bibnamefont
  {Torlina}}, \bibinfo {author} {\bibfnamefont {F.}~\bibnamefont {Morales}},
  \bibinfo {author} {\bibfnamefont {J.}~\bibnamefont {Kaushal}}, \bibinfo
  {author} {\bibfnamefont {I.}~\bibnamefont {Ivanov}}, \bibinfo {author}
  {\bibfnamefont {A.}~\bibnamefont {Kheifets}}, \bibinfo {author}
  {\bibfnamefont {A.}~\bibnamefont {Zielinski}}, \bibinfo {author}
  {\bibfnamefont {A.}~\bibnamefont {Scrinzi}}, \bibinfo {author} {\bibfnamefont
  {H.~G.}\ \bibnamefont {Muller}}, \bibinfo {author} {\bibfnamefont
  {S.}~\bibnamefont {Sukiasyan}}, \bibinfo {author} {\bibfnamefont
  {M.}~\bibnamefont {Ivanov}}, \ and\ \bibinfo {author} {\bibfnamefont
  {O.}~\bibnamefont {Smirnova}},\ }\href {\doibase
  http://dx.doi.org/10.1038/nphys3340} {\bibfield  {journal} {\bibinfo
  {journal} {Nature Physics}\ }\textbf {\bibinfo {volume} {11}},\ \bibinfo
  {pages} {503} (\bibinfo {year} {2015})}\BibitemShut {NoStop}%
\bibitem [{\citenamefont {Pedatzur}\ \emph {et~al.}(2015)\citenamefont
  {Pedatzur}, \citenamefont {Orenstein}, \citenamefont {Serbinenko},
  \citenamefont {Soifer}, \citenamefont {Bruner}, \citenamefont {Uzan},
  \citenamefont {Brambila}, \citenamefont {Harvey}, \citenamefont {Torlina},
  \citenamefont {Morales}, \citenamefont {Smirnova},\ and\ \citenamefont
  {Dudovich}}]{POS15}%
  \BibitemOpen
  \bibfield  {author} {\bibinfo {author} {\bibfnamefont {O.}~\bibnamefont
  {Pedatzur}}, \bibinfo {author} {\bibfnamefont {G.}~\bibnamefont {Orenstein}},
  \bibinfo {author} {\bibfnamefont {V.}~\bibnamefont {Serbinenko}}, \bibinfo
  {author} {\bibfnamefont {H.}~\bibnamefont {Soifer}}, \bibinfo {author}
  {\bibfnamefont {B.~D.}\ \bibnamefont {Bruner}}, \bibinfo {author}
  {\bibfnamefont {A.~J.}\ \bibnamefont {Uzan}}, \bibinfo {author}
  {\bibfnamefont {D.~S.}\ \bibnamefont {Brambila}}, \bibinfo {author}
  {\bibfnamefont {A.~G.}\ \bibnamefont {Harvey}}, \bibinfo {author}
  {\bibfnamefont {L.}~\bibnamefont {Torlina}}, \bibinfo {author} {\bibfnamefont
  {F.}~\bibnamefont {Morales}}, \bibinfo {author} {\bibfnamefont
  {O.}~\bibnamefont {Smirnova}}, \ and\ \bibinfo {author} {\bibfnamefont
  {N.}~\bibnamefont {Dudovich}},\ }\href {\doibase
  http://dx.doi.org/10.1038/nphys3436} {\bibfield  {journal} {\bibinfo
  {journal} {Nature Physics}\ }\textbf {\bibinfo {volume} {11}},\ \bibinfo
  {pages} {815} (\bibinfo {year} {2015})}\BibitemShut {NoStop}%
\bibitem [{\citenamefont {Zimmermann}\ \emph {et~al.}(2016)\citenamefont
  {Zimmermann}, \citenamefont {Mishra}, \citenamefont {Doran}, \citenamefont
  {Gordon},\ and\ \citenamefont {Landsman}}]{ZMD16}%
  \BibitemOpen
  \bibfield  {author} {\bibinfo {author} {\bibfnamefont {T.}~\bibnamefont
  {Zimmermann}}, \bibinfo {author} {\bibfnamefont {S.}~\bibnamefont {Mishra}},
  \bibinfo {author} {\bibfnamefont {B.~R.}\ \bibnamefont {Doran}}, \bibinfo
  {author} {\bibfnamefont {D.~F.}\ \bibnamefont {Gordon}}, \ and\ \bibinfo
  {author} {\bibfnamefont {A.~S.}\ \bibnamefont {Landsman}},\ }\href {\doibase
  10.1103/PhysRevLett.116.233603} {\bibfield  {journal} {\bibinfo  {journal}
  {Phys. Rev. Lett.}\ }\textbf {\bibinfo {volume} {116}},\ \bibinfo {pages}
  {233603} (\bibinfo {year} {2016})}\BibitemShut {NoStop}%
\bibitem [{\citenamefont {Chiao}\ \emph {et~al.}(1991)\citenamefont {Chiao},
  \citenamefont {Kwiat},\ and\ \citenamefont {Steinberg}}]{CKS91}%
  \BibitemOpen
  \bibfield  {author} {\bibinfo {author} {\bibfnamefont {R.}~\bibnamefont
  {Chiao}}, \bibinfo {author} {\bibfnamefont {P.}~\bibnamefont {Kwiat}}, \ and\
  \bibinfo {author} {\bibfnamefont {A.}~\bibnamefont {Steinberg}},\ }\href
  {\doibase http://dx.doi.org/10.1016/0921-4526(91)90724-S} {\bibfield
  {journal} {\bibinfo  {journal} {Physica B: Condensed Matter}\ }\textbf
  {\bibinfo {volume} {175}},\ \bibinfo {pages} {257 } (\bibinfo {year}
  {1991})}\BibitemShut {NoStop}%
\bibitem [{\citenamefont {Steinberg}\ \emph {et~al.}(1993)\citenamefont
  {Steinberg}, \citenamefont {Kwiat},\ and\ \citenamefont {Chiao}}]{SKC93}%
  \BibitemOpen
  \bibfield  {author} {\bibinfo {author} {\bibfnamefont {A.~M.}\ \bibnamefont
  {Steinberg}}, \bibinfo {author} {\bibfnamefont {P.~G.}\ \bibnamefont
  {Kwiat}}, \ and\ \bibinfo {author} {\bibfnamefont {R.~Y.}\ \bibnamefont
  {Chiao}},\ }\href {\doibase 10.1103/PhysRevLett.71.708} {\bibfield  {journal}
  {\bibinfo  {journal} {Phys. Rev. Lett.}\ }\textbf {\bibinfo {volume} {71}},\
  \bibinfo {pages} {708} (\bibinfo {year} {1993})}\BibitemShut {NoStop}%
\bibitem [{\citenamefont {C.R.}\ and\ \citenamefont {R.}()}]{LSM98}%
  \BibitemOpen
  \bibfield  {author} {\bibinfo {author} {\bibfnamefont {L.}~\bibnamefont
  {C.R.}}\ and\ \bibinfo {author} {\bibfnamefont {S.~M.}\ \bibnamefont {R.}},\
  }\href {\doibase
  10.1002/(SICI)1521-3889(199812)7:7/8<662::AID-ANDP662>3.0.CO;2-T} {\bibfield
  {journal} {\bibinfo  {journal} {Annalen der Physik}\ }\textbf {\bibinfo
  {volume} {7}},\ \bibinfo {pages} {662}}\BibitemShut {NoStop}%
\bibitem [{\citenamefont {Krekora}\ \emph {et~al.}(2001)\citenamefont
  {Krekora}, \citenamefont {Su},\ and\ \citenamefont {Grobe}}]{KSG01}%
  \BibitemOpen
  \bibfield  {author} {\bibinfo {author} {\bibfnamefont {P.}~\bibnamefont
  {Krekora}}, \bibinfo {author} {\bibfnamefont {Q.}~\bibnamefont {Su}}, \ and\
  \bibinfo {author} {\bibfnamefont {R.}~\bibnamefont {Grobe}},\ }\href
  {\doibase 10.1103/PhysRevA.63.032107} {\bibfield  {journal} {\bibinfo
  {journal} {Phys. Rev. A}\ }\textbf {\bibinfo {volume} {63}},\ \bibinfo
  {pages} {032107} (\bibinfo {year} {2001})}\BibitemShut {NoStop}%
\bibitem [{\citenamefont {C‐F.}\ and\ \citenamefont {X.}()}]{LCh02}%
  \BibitemOpen
  \bibfield  {author} {\bibinfo {author} {\bibfnamefont {L.}~\bibnamefont
  {C‐F.}}\ and\ \bibinfo {author} {\bibfnamefont {C.}~\bibnamefont {X.}},\
  }\href {\doibase 10.1002/1521-3889(200212)11:12<916::AID-ANDP916>3.0.CO;2-V}
  {\bibfield  {journal} {\bibinfo  {journal} {Annalen der Physik}\ }\textbf
  {\bibinfo {volume} {11}},\ \bibinfo {pages} {916}}\BibitemShut {NoStop}%
\bibitem [{\citenamefont {Petrillo}\ and\ \citenamefont
  {Janner}(2003)}]{PJa03}%
  \BibitemOpen
  \bibfield  {author} {\bibinfo {author} {\bibfnamefont {V.}~\bibnamefont
  {Petrillo}}\ and\ \bibinfo {author} {\bibfnamefont {D.}~\bibnamefont
  {Janner}},\ }\href {\doibase 10.1103/PhysRevA.67.012110} {\bibfield
  {journal} {\bibinfo  {journal} {Phys. Rev. A}\ }\textbf {\bibinfo {volume}
  {67}},\ \bibinfo {pages} {012110} (\bibinfo {year} {2003})}\BibitemShut
  {NoStop}%
\bibitem [{\citenamefont {Chen}\ and\ \citenamefont {Li}(2003)}]{CLi03}%
  \BibitemOpen
  \bibfield  {author} {\bibinfo {author} {\bibfnamefont {X.}~\bibnamefont
  {Chen}}\ and\ \bibinfo {author} {\bibfnamefont {C.-F.}\ \bibnamefont {Li}},\
  }\href {\doibase 10.1103/PhysRevA.68.052105} {\bibfield  {journal} {\bibinfo
  {journal} {Phys. Rev. A}\ }\textbf {\bibinfo {volume} {68}},\ \bibinfo
  {pages} {052105} (\bibinfo {year} {2003})}\BibitemShut {NoStop}%
\bibitem [{\citenamefont {De~Leo}\ and\ \citenamefont {Rotelli}(2007)}]{LRo07}%
  \BibitemOpen
  \bibfield  {author} {\bibinfo {author} {\bibfnamefont {S.}~\bibnamefont
  {De~Leo}}\ and\ \bibinfo {author} {\bibfnamefont {P.}~\bibnamefont
  {Rotelli}},\ }\href {\doibase 10.1140/epjc/s10052-007-0297-4} {\bibfield
  {journal} {\bibinfo  {journal} {The European Physical Journal C}\ }\textbf
  {\bibinfo {volume} {51}},\ \bibinfo {pages} {241} (\bibinfo {year}
  {2007})}\BibitemShut {NoStop}%
\bibitem [{\citenamefont {Lunardi}\ and\ \citenamefont
  {Manzoni}(2007)}]{LMa07}%
  \BibitemOpen
  \bibfield  {author} {\bibinfo {author} {\bibfnamefont {J.~T.}\ \bibnamefont
  {Lunardi}}\ and\ \bibinfo {author} {\bibfnamefont {L.~A.}\ \bibnamefont
  {Manzoni}},\ }\href {\doibase http://dx.doi.org/10.1103/PhysRevA.76.042111}
  {\bibfield  {journal} {\bibinfo  {journal} {Physical Review A}\ }\textbf
  {\bibinfo {volume} {76}},\ \bibinfo {pages} {042111} (\bibinfo {year}
  {2007})}\BibitemShut {NoStop}%
\bibitem [{\citenamefont {Lunardi}\ \emph
  {et~al.}(2011{\natexlab{a}})\citenamefont {Lunardi}, \citenamefont {Manzoni},
  \citenamefont {Nystrom},\ and\ \citenamefont {Perreault}}]{LMN11b}%
  \BibitemOpen
  \bibfield  {author} {\bibinfo {author} {\bibfnamefont {J.~T.}\ \bibnamefont
  {Lunardi}}, \bibinfo {author} {\bibfnamefont {L.~A.}\ \bibnamefont
  {Manzoni}}, \bibinfo {author} {\bibfnamefont {A.~T.}\ \bibnamefont
  {Nystrom}}, \ and\ \bibinfo {author} {\bibfnamefont {B.~M.}\ \bibnamefont
  {Perreault}},\ }\href {\doibase 10.1007/s10946-011-9232-0} {\bibfield
  {journal} {\bibinfo  {journal} {Journal of Russian Laser Research}\ }\textbf
  {\bibinfo {volume} {32}},\ \bibinfo {pages} {431} (\bibinfo {year}
  {2011}{\natexlab{a}})}\BibitemShut {NoStop}%
\bibitem [{\citenamefont {Peres}(1980)}]{Per80}%
  \BibitemOpen
  \bibfield  {author} {\bibinfo {author} {\bibfnamefont {A.}~\bibnamefont
  {Peres}},\ }\href {\doibase http://dx.doi.org/10.1119/1.12061} {\bibfield
  {journal} {\bibinfo  {journal} {American Journal of Physics}\ }\textbf
  {\bibinfo {volume} {48}},\ \bibinfo {pages} {552} (\bibinfo {year}
  {1980})}\BibitemShut {NoStop}%
\bibitem [{\citenamefont {Wigner}(1955)}]{Wig55}%
  \BibitemOpen
  \bibfield  {author} {\bibinfo {author} {\bibfnamefont {E.~P.}\ \bibnamefont
  {Wigner}},\ }\href {\doibase http://dx.doi.org/10.1103/PhysRev.98.145}
  {\bibfield  {journal} {\bibinfo  {journal} {Physical Review}\ }\textbf
  {\bibinfo {volume} {98}},\ \bibinfo {pages} {145} (\bibinfo {year}
  {1955})}\BibitemShut {NoStop}%
\bibitem [{\citenamefont {Smith}(1960)}]{Smi60}%
  \BibitemOpen
  \bibfield  {author} {\bibinfo {author} {\bibfnamefont {F.~T.}\ \bibnamefont
  {Smith}},\ }\href {\doibase http://dx.doi.org/10.1103/PhysRev.118.349}
  {\bibfield  {journal} {\bibinfo  {journal} {Physical Review}\ }\textbf
  {\bibinfo {volume} {118}},\ \bibinfo {pages} {349} (\bibinfo {year}
  {1960})}\BibitemShut {NoStop}%
\bibitem [{\citenamefont {Baz'}(1967)}]{Baz67}%
  \BibitemOpen
  \bibfield  {author} {\bibinfo {author} {\bibfnamefont {A.~I.}\ \bibnamefont
  {Baz'}},\ }\href@noop {} {\bibfield  {journal} {\bibinfo  {journal} {Soviet
  Journal of Nuclear Physics}\ }\textbf {\bibinfo {volume} {4}},\ \bibinfo
  {pages} {182} (\bibinfo {year} {1967})}\BibitemShut {NoStop}%
\bibitem [{\citenamefont {Rybachenko}(1967)}]{Ryb67}%
  \BibitemOpen
  \bibfield  {author} {\bibinfo {author} {\bibfnamefont {V.~F.}\ \bibnamefont
  {Rybachenko}},\ }\href@noop {} {\bibfield  {journal} {\bibinfo  {journal}
  {Soviet Journal of Nuclear Physics}\ }\textbf {\bibinfo {volume} {5}},\
  \bibinfo {pages} {635} (\bibinfo {year} {1967})}\BibitemShut {NoStop}%
\bibitem [{\citenamefont {B\"uttiker}(1983)}]{But83}%
  \BibitemOpen
  \bibfield  {author} {\bibinfo {author} {\bibfnamefont {M.}~\bibnamefont
  {B\"uttiker}},\ }\href {\doibase http://dx.doi.org/10.1103/PhysRevB.27.6178}
  {\bibfield  {journal} {\bibinfo  {journal} {Physical Review B}\ }\textbf
  {\bibinfo {volume} {27}},\ \bibinfo {pages} {6178} (\bibinfo {year}
  {1983})}\BibitemShut {NoStop}%
\bibitem [{\citenamefont {Falck}\ and\ \citenamefont {Hauge}(1988)}]{FHa88}%
  \BibitemOpen
  \bibfield  {author} {\bibinfo {author} {\bibfnamefont {J.~P.}\ \bibnamefont
  {Falck}}\ and\ \bibinfo {author} {\bibfnamefont {E.~H.}\ \bibnamefont
  {Hauge}},\ }\href {\doibase http://dx.doi.org/10.1103/PhysRevB.38.3287}
  {\bibfield  {journal} {\bibinfo  {journal} {Physical Review B}\ }\textbf
  {\bibinfo {volume} {38}},\ \bibinfo {pages} {3287} (\bibinfo {year}
  {1988})}\BibitemShut {NoStop}%
\bibitem [{\citenamefont {Salecker}\ and\ \citenamefont
  {Wigner}(1958)}]{SWi58}%
  \BibitemOpen
  \bibfield  {author} {\bibinfo {author} {\bibfnamefont {H.}~\bibnamefont
  {Salecker}}\ and\ \bibinfo {author} {\bibfnamefont {E.~P.}\ \bibnamefont
  {Wigner}},\ }\href {\doibase http://dx.doi.org/10.1103/PhysRev.109.571}
  {\bibfield  {journal} {\bibinfo  {journal} {Physical Review}\ }\textbf
  {\bibinfo {volume} {109}},\ \bibinfo {pages} {571} (\bibinfo {year}
  {1958})}\BibitemShut {NoStop}%
\bibitem [{\citenamefont {Eckle}\ \emph {et~al.}(2008)\citenamefont {Eckle},
  \citenamefont {Pfeiffer}, \citenamefont {Cirelli}, \citenamefont {Staudte},
  \citenamefont {D{\"o}rner}, \citenamefont {Muller}, \citenamefont
  {B{\"u}ttiker},\ and\ \citenamefont {Keller}}]{EPC08}%
  \BibitemOpen
  \bibfield  {author} {\bibinfo {author} {\bibfnamefont {P.}~\bibnamefont
  {Eckle}}, \bibinfo {author} {\bibfnamefont {A.~N.}\ \bibnamefont {Pfeiffer}},
  \bibinfo {author} {\bibfnamefont {C.}~\bibnamefont {Cirelli}}, \bibinfo
  {author} {\bibfnamefont {A.}~\bibnamefont {Staudte}}, \bibinfo {author}
  {\bibfnamefont {R.}~\bibnamefont {D{\"o}rner}}, \bibinfo {author}
  {\bibfnamefont {H.~G.}\ \bibnamefont {Muller}}, \bibinfo {author}
  {\bibfnamefont {M.}~\bibnamefont {B{\"u}ttiker}}, \ and\ \bibinfo {author}
  {\bibfnamefont {U.}~\bibnamefont {Keller}},\ }\href {\doibase
  10.1126/science.1163439} {\bibfield  {journal} {\bibinfo  {journal}
  {Science}\ }\textbf {\bibinfo {volume} {322}},\ \bibinfo {pages} {1525}
  (\bibinfo {year} {2008})}\BibitemShut {NoStop}%
\bibitem [{\citenamefont {Orlando}\ \emph {et~al.}(2014)\citenamefont
  {Orlando}, \citenamefont {McDonald}, \citenamefont {Protik}, \citenamefont
  {Vampa},\ and\ \citenamefont {Brabec}}]{OMP14}%
  \BibitemOpen
  \bibfield  {author} {\bibinfo {author} {\bibfnamefont {G.}~\bibnamefont
  {Orlando}}, \bibinfo {author} {\bibfnamefont {C.~R.}\ \bibnamefont
  {McDonald}}, \bibinfo {author} {\bibfnamefont {N.~H.}\ \bibnamefont
  {Protik}}, \bibinfo {author} {\bibfnamefont {G.}~\bibnamefont {Vampa}}, \
  and\ \bibinfo {author} {\bibfnamefont {T.}~\bibnamefont {Brabec}},\ }\href
  {http://stacks.iop.org/0953-4075/47/i=20/a=204002} {\bibfield  {journal}
  {\bibinfo  {journal} {Journal of Physics B: Atomic, Molecular and Optical
  Physics}\ }\textbf {\bibinfo {volume} {47}},\ \bibinfo {pages} {204002}
  (\bibinfo {year} {2014})}\BibitemShut {NoStop}%
\bibitem [{\citenamefont {Landsman}\ and\ \citenamefont
  {Keller}(2015)}]{LKe15}%
  \BibitemOpen
  \bibfield  {author} {\bibinfo {author} {\bibfnamefont {A.~S.}\ \bibnamefont
  {Landsman}}\ and\ \bibinfo {author} {\bibfnamefont {U.}~\bibnamefont
  {Keller}},\ }\href {\doibase http://dx.doi.org/10.1016/j.physrep.2014.09.002}
  {\bibfield  {journal} {\bibinfo  {journal} {Physics Reports}\ }\textbf
  {\bibinfo {volume} {547}},\ \bibinfo {pages} {1} (\bibinfo {year}
  {2015})}\BibitemShut {NoStop}%
\bibitem [{\citenamefont {Lunardi}\ \emph
  {et~al.}(2011{\natexlab{b}})\citenamefont {Lunardi}, \citenamefont
  {Manzoni},\ and\ \citenamefont {Nystrom}}]{LMN11a}%
  \BibitemOpen
  \bibfield  {author} {\bibinfo {author} {\bibfnamefont {J.~T.}\ \bibnamefont
  {Lunardi}}, \bibinfo {author} {\bibfnamefont {L.~A.}\ \bibnamefont
  {Manzoni}}, \ and\ \bibinfo {author} {\bibfnamefont {A.~T.}\ \bibnamefont
  {Nystrom}},\ }\href {\doibase
  http://dx.doi.org/10.1016/j.physleta.2010.11.055} {\bibfield  {journal}
  {\bibinfo  {journal} {Physics Letters A}\ }\textbf {\bibinfo {volume}
  {375}},\ \bibinfo {pages} {415} (\bibinfo {year}
  {2011}{\natexlab{b}})}\BibitemShut {NoStop}%
\bibitem [{\citenamefont {Kelkar}\ \emph {et~al.}(2009)\citenamefont {Kelkar},
  \citenamefont {Casta{\~n}eda},\ and\ \citenamefont {Nowakowski}}]{KCN09}%
  \BibitemOpen
  \bibfield  {author} {\bibinfo {author} {\bibfnamefont {N.~G.}\ \bibnamefont
  {Kelkar}}, \bibinfo {author} {\bibfnamefont {H.~M.}\ \bibnamefont
  {Casta{\~n}eda}}, \ and\ \bibinfo {author} {\bibfnamefont {M.}~\bibnamefont
  {Nowakowski}},\ }\href {http://stacks.iop.org/0295-5075/85/i=2/a=20006}
  {\bibfield  {journal} {\bibinfo  {journal} {EPL (Europhysics Letters)}\
  }\textbf {\bibinfo {volume} {85}},\ \bibinfo {pages} {20006} (\bibinfo {year}
  {2009})}\BibitemShut {NoStop}%
\bibitem [{\citenamefont {Goto}\ \emph {et~al.}(2004)\citenamefont {Goto},
  \citenamefont {Iwamoto}, \citenamefont {de~Aquino}, \citenamefont
  {Aguilera-Navarro},\ and\ \citenamefont {Kobe}}]{GIA04}%
  \BibitemOpen
  \bibfield  {author} {\bibinfo {author} {\bibfnamefont {M.}~\bibnamefont
  {Goto}}, \bibinfo {author} {\bibfnamefont {H.}~\bibnamefont {Iwamoto}},
  \bibinfo {author} {\bibfnamefont {V.~M.}\ \bibnamefont {de~Aquino}}, \bibinfo
  {author} {\bibfnamefont {V.~C.}\ \bibnamefont {Aguilera-Navarro}}, \ and\
  \bibinfo {author} {\bibfnamefont {D.~H.}\ \bibnamefont {Kobe}},\ }\href
  {http://stacks.iop.org/0305-4470/37/i=11/a=005} {\bibfield  {journal}
  {\bibinfo  {journal} {Journal of Physics A: Mathematical and General}\
  }\textbf {\bibinfo {volume} {37}},\ \bibinfo {pages} {3599} (\bibinfo {year}
  {2004})}\BibitemShut {NoStop}%
\bibitem [{\citenamefont {Brouard}\ \emph {et~al.}(1994)\citenamefont
  {Brouard}, \citenamefont {Sala},\ and\ \citenamefont {Muga}}]{BSM94}%
  \BibitemOpen
  \bibfield  {author} {\bibinfo {author} {\bibfnamefont {S.}~\bibnamefont
  {Brouard}}, \bibinfo {author} {\bibfnamefont {R.}~\bibnamefont {Sala}}, \
  and\ \bibinfo {author} {\bibfnamefont {J.~G.}\ \bibnamefont {Muga}},\ }\href
  {\doibase 10.1103/PhysRevA.49.4312} {\bibfield  {journal} {\bibinfo
  {journal} {Phys. Rev. A}\ }\textbf {\bibinfo {volume} {49}},\ \bibinfo
  {pages} {4312} (\bibinfo {year} {1994})}\BibitemShut {NoStop}%
\bibitem [{\citenamefont {Petrillo}\ and\ \citenamefont
  {Olkhovsky}(2006)}]{POl06}%
  \BibitemOpen
  \bibfield  {author} {\bibinfo {author} {\bibfnamefont {V.}~\bibnamefont
  {Petrillo}}\ and\ \bibinfo {author} {\bibfnamefont {V.~S.}\ \bibnamefont
  {Olkhovsky}},\ }\href {http://stacks.iop.org/0295-5075/74/i=2/a=327}
  {\bibfield  {journal} {\bibinfo  {journal} {EPL (Europhysics Letters)}\
  }\textbf {\bibinfo {volume} {74}},\ \bibinfo {pages} {327} (\bibinfo {year}
  {2006})}\BibitemShut {NoStop}%
\bibitem [{\citenamefont {Hartman}(1962)}]{Har62}%
  \BibitemOpen
  \bibfield  {author} {\bibinfo {author} {\bibfnamefont {T.~E.}\ \bibnamefont
  {Hartman}},\ }\href {\doibase http://dx.doi.org/10.1063/1.1702424} {\bibfield
   {journal} {\bibinfo  {journal} {Journal of Applied Physics}\ }\textbf
  {\bibinfo {volume} {33}},\ \bibinfo {pages} {3427} (\bibinfo {year}
  {1962})}\BibitemShut {NoStop}%
\bibitem [{\citenamefont {Frentz}\ \emph {et~al.}(2014)\citenamefont {Frentz},
  \citenamefont {Lunardi},\ and\ \citenamefont {Manzoni}}]{FLM14}%
  \BibitemOpen
  \bibfield  {author} {\bibinfo {author} {\bibfnamefont {B.~A.}\ \bibnamefont
  {Frentz}}, \bibinfo {author} {\bibfnamefont {J.~T.}\ \bibnamefont {Lunardi}},
  \ and\ \bibinfo {author} {\bibfnamefont {L.~A.}\ \bibnamefont {Manzoni}},\
  }\href {\doibase http://dx.doi.org/10.1140/epjp/i2014-14005-7} {\bibfield
  {journal} {\bibinfo  {journal} {European Physical Journal Plus}\ }\textbf
  {\bibinfo {volume} {129}},\ \bibinfo {pages} {5} (\bibinfo {year}
  {2014})}\BibitemShut {NoStop}%
\bibitem [{\citenamefont {Turok}(2014)}]{Tur14}%
  \BibitemOpen
  \bibfield  {author} {\bibinfo {author} {\bibfnamefont {N.}~\bibnamefont
  {Turok}},\ }\href {http://stacks.iop.org/1367-2630/16/i=6/a=063006}
  {\bibfield  {journal} {\bibinfo  {journal} {New Journal of Physics}\ }\textbf
  {\bibinfo {volume} {16}},\ \bibinfo {pages} {063006} (\bibinfo {year}
  {2014})}\BibitemShut {NoStop}%
\bibitem [{\citenamefont {Ban}\ \emph {et~al.}(2010)\citenamefont {Ban},
  \citenamefont {Sherman}, \citenamefont {Muga},\ and\ \citenamefont
  {B\"uttiker}}]{BSM10}%
  \BibitemOpen
  \bibfield  {author} {\bibinfo {author} {\bibfnamefont {Y.}~\bibnamefont
  {Ban}}, \bibinfo {author} {\bibfnamefont {E.~Y.}\ \bibnamefont {Sherman}},
  \bibinfo {author} {\bibfnamefont {J.~G.}\ \bibnamefont {Muga}}, \ and\
  \bibinfo {author} {\bibfnamefont {M.}~\bibnamefont {B\"uttiker}},\ }\href
  {\doibase 10.1103/PhysRevA.82.062121} {\bibfield  {journal} {\bibinfo
  {journal} {Phys. Rev. A}\ }\textbf {\bibinfo {volume} {82}},\ \bibinfo
  {pages} {062121} (\bibinfo {year} {2010})}\BibitemShut {NoStop}%
\bibitem [{\citenamefont {Cal\c{c}ada}\ \emph {et~al.}(2009)\citenamefont
  {Cal\c{c}ada}, \citenamefont {Lunardi},\ and\ \citenamefont
  {Manzoni}}]{CLM09}%
  \BibitemOpen
  \bibfield  {author} {\bibinfo {author} {\bibfnamefont {M.}~\bibnamefont
  {Cal\c{c}ada}}, \bibinfo {author} {\bibfnamefont {J.~T.}\ \bibnamefont
  {Lunardi}}, \ and\ \bibinfo {author} {\bibfnamefont {L.~A.}\ \bibnamefont
  {Manzoni}},\ }\href {\doibase http://dx.doi.org/10.1103/PhysRevA.79.012110}
  {\bibfield  {journal} {\bibinfo  {journal} {Physical Review A}\ }\textbf
  {\bibinfo {volume} {79}},\ \bibinfo {pages} {012110} (\bibinfo {year}
  {2009})}\BibitemShut {NoStop}%
\bibitem [{Note1()}]{Note1}%
  \BibitemOpen
  \bibinfo {note} {Specifically, we use the fact that \\ $$\delta (g(x)) =
  \DOTSB \sum@ \slimits@ _j \protect \frac {\delta (x-x_j)}{\left | g^\prime
  \left ( x_j\right ) \right | },$$ \\ where $\protect \{ x_j \protect \}$ is
  the set of zeros of the function $g(x)$ and the prime indicates a derivative
  with respect to the independent variable.}\BibitemShut {Stop}%
\bibitem [{\citenamefont {Park}(2011)}]{Par11}%
  \BibitemOpen
  \bibfield  {author} {\bibinfo {author} {\bibfnamefont {C.-S.}\ \bibnamefont
  {Park}},\ }\href {\doibase http://dx.doi.org/10.1016/j.physleta.2011.07.048}
  {\bibfield  {journal} {\bibinfo  {journal} {Physics Letters A}\ }\textbf
  {\bibinfo {volume} {375}},\ \bibinfo {pages} {3348 } (\bibinfo {year}
  {2011})}\BibitemShut {NoStop}%
\bibitem [{\citenamefont {Park}(2009)}]{Par09}%
  \BibitemOpen
  \bibfield  {author} {\bibinfo {author} {\bibfnamefont {C.-S.}\ \bibnamefont
  {Park}},\ }\href {\doibase http://dx.doi.org/10.1103/PhysRevA.80.012111}
  {\bibfield  {journal} {\bibinfo  {journal} {Physical Review A}\ }\textbf
  {\bibinfo {volume} {80}},\ \bibinfo {pages} {012111} (\bibinfo {year}
  {2009})}\BibitemShut {NoStop}%
\bibitem [{\citenamefont {Aharonov}\ \emph {et~al.}(2002)\citenamefont
  {Aharonov}, \citenamefont {Erez},\ and\ \citenamefont {Reznik}}]{AER02}%
  \BibitemOpen
  \bibfield  {author} {\bibinfo {author} {\bibfnamefont {Y.}~\bibnamefont
  {Aharonov}}, \bibinfo {author} {\bibfnamefont {N.}~\bibnamefont {Erez}}, \
  and\ \bibinfo {author} {\bibfnamefont {B.}~\bibnamefont {Reznik}},\ }\href
  {\doibase http://dx.doi.org/10.1103/PhysRevA.65.052124} {\bibfield  {journal}
  {\bibinfo  {journal} {Physical Review A}\ }\textbf {\bibinfo {volume} {65}},\
  \bibinfo {pages} {052124} (\bibinfo {year} {2002})}\BibitemShut {NoStop}%
\bibitem [{\citenamefont {Aharonov}\ \emph {et~al.}(2003)\citenamefont
  {Aharonov}, \citenamefont {Erez},\ and\ \citenamefont {Reznik}}]{AER03}%
  \BibitemOpen
  \bibfield  {author} {\bibinfo {author} {\bibfnamefont {Y.}~\bibnamefont
  {Aharonov}}, \bibinfo {author} {\bibfnamefont {N.}~\bibnamefont {Erez}}, \
  and\ \bibinfo {author} {\bibfnamefont {B.}~\bibnamefont {Reznik}},\ }\href
  {\doibase http://dx.doi.org/10.1080/09500340308234558} {\bibfield  {journal}
  {\bibinfo  {journal} {Journal of Modern Optics}\ }\textbf {\bibinfo {volume}
  {50}},\ \bibinfo {pages} {1139} (\bibinfo {year} {2003})}\BibitemShut
  {NoStop}%
\bibitem [{\citenamefont {Lee}\ \emph {et~al.}(2016)\citenamefont {Lee},
  \citenamefont {Lunardi}, \citenamefont {Manzoni},\ and\ \citenamefont
  {Nyquist}}]{LLM16}%
  \BibitemOpen
  \bibfield  {author} {\bibinfo {author} {\bibfnamefont {M.~A.}\ \bibnamefont
  {Lee}}, \bibinfo {author} {\bibfnamefont {J.~T.}\ \bibnamefont {Lunardi}},
  \bibinfo {author} {\bibfnamefont {L.~A.}\ \bibnamefont {Manzoni}}, \ and\
  \bibinfo {author} {\bibfnamefont {E.~A.}\ \bibnamefont {Nyquist}},\ }\href
  {\doibase 10.3389/fphy.2016.00010} {\bibfield  {journal} {\bibinfo  {journal}
  {Frontiers in Physics}\ }\textbf {\bibinfo {volume} {4}},\ \bibinfo {pages}
  {10} (\bibinfo {year} {2016})}\BibitemShut {NoStop}%
\bibitem [{\citenamefont {Teeny}\ \emph {et~al.}(2016)\citenamefont {Teeny},
  \citenamefont {Yakaboylu}, \citenamefont {Bauke},\ and\ \citenamefont
  {Keitel}}]{TYB16}%
  \BibitemOpen
  \bibfield  {author} {\bibinfo {author} {\bibfnamefont {N.}~\bibnamefont
  {Teeny}}, \bibinfo {author} {\bibfnamefont {E.}~\bibnamefont {Yakaboylu}},
  \bibinfo {author} {\bibfnamefont {H.}~\bibnamefont {Bauke}}, \ and\ \bibinfo
  {author} {\bibfnamefont {C.~H.}\ \bibnamefont {Keitel}},\ }\href {\doibase
  10.1103/PhysRevLett.116.063003} {\bibfield  {journal} {\bibinfo  {journal}
  {Phys. Rev. Lett.}\ }\textbf {\bibinfo {volume} {116}},\ \bibinfo {pages}
  {063003} (\bibinfo {year} {2016})}\BibitemShut {NoStop}%
\end{thebibliography}%


\end{document}